\documentclass[english,pccp,preprint,endfloats,superscriptaddress]{revtex4-1} 
\usepackage{epsfig}
\usepackage{amsmath} 
\usepackage{graphicx} 
\usepackage{longtable} 
\usepackage{hyperref} 
\usepackage{amssymb} 
\usepackage{dcolumn} 
\usepackage{mhchem} 
\usepackage{cleveref} 
\usepackage{subfigure} 
\usepackage{array} 
\usepackage{multirow} 
\usepackage{tabularx} 
\usepackage{cleveref} 
\DeclareMathOperator\erf{erf} 
\DeclareMathOperator\erfc{erfc} 

\begin{document} 

\title{Van der Waals interactions in rare-gas dimers: The role of interparticle interactions} 

\author{Yu-Ting Chen} 
\affiliation{Department of Physics, National Taiwan University, Taipei 10617, Taiwan} 

\author{Kerwin Hui} 
\affiliation{Department of Physics, National Taiwan University, Taipei 10617, Taiwan} 

\author{Jeng-Da Chai} 
\email[Author to whom correspondence should be addressed. Electronic mail: ]{jdchai@phys.ntu.edu.tw} 
\affiliation{Department of Physics, National Taiwan University, Taipei 10617, Taiwan} 
\affiliation{Center for Theoretical Sciences and Center for Quantum Science and Engineering, National Taiwan University, Taipei 10617, Taiwan} 

\date{\today} 

\begin{abstract} 

We investigate the potential energy curves of rare-gas dimers with various ranges and strengths of interparticle interactions (nuclear-electron, electron-electron, and nuclear-nuclear interactions). Our investigation is based on 
the highly accurate coupled-cluster theory associated with those interparticle interactions. For comparison, the performance of the corresponding Hartree-Fock theory, second-order M\o ller-Plesset perturbation theory, and 
density functional theory is also investigated. Our results reveal that when the interparticle interactions retain the long-range Coulomb tails, the nature of van der Waals interactions in the rare-gas dimers remains similar. By 
contrast, when the interparticle interactions are sufficiently short-range, the conventional van der Waals interactions in the rare-gas dimers completely disappear, yielding purely repulsive potential energy curves. 

\end{abstract} 

\maketitle

\section{Introduction} 

Van der Waals (vdW) interactions \cite{vdW,vdW2,Jensen,MM,n1} are omnipresent in materials and biological systems. These interactions are of fundamental importance in numerous fields, involving molecular and condensed 
matter physics, supramolecular chemistry, structural biology, surface science, and nanoscience. While vdW interactions are individually weak (e.g., compared to covalent bonds or electrostatic interactions between permanent 
charges, dipoles, etc.), they are collectively important in the determination of the structure, stability, and function of a vast variety of systems, such as the interaction between graphene layers, the self-assembly of functional 
nanomaterials, the structure of biomacromolecules (e.g., DNA, RNA, and proteins), and the molecular recognition of proteins \cite{SA}. 

In particular, the potential energy curve of a rare-gas dimer is predominantly determined by the interplay between the exchange-repulsion energy at short internuclear distances and the attractive vdW interaction at large internuclear 
distances, exhibiting a potential minimum (the vdW minimum) at an intermediate internuclear distance. The exchange-repulsion energy arises from the overlap of the electron densities of the two atoms. On the other hand, the vdW 
interaction, also known as London dispersion interaction or induced dipole-induced dipole interaction, arises from the Coulomb correlation of electron density fluctuations in the two well-separated atoms. The potential energy curve 
can be conveniently approximated by the Lennard-Jones (LJ) potential \cite{Jensen} 
\begin{equation} 
V_{\text{LJ}}(R) = 4 \epsilon \bigg[\bigg(\frac{\sigma}{R}\bigg)^{12} - \bigg(\frac{\sigma}{R}\bigg)^{6}\bigg], 
\label{eq:LJ} 
\end{equation} 
where $R$ is the internuclear distance, $\sigma$ is the distance at which the potential is zero, and $-\epsilon$ is the minimum of the potential, which is reached at $R = 2^{1/6} \sigma$. Here the $R^{-12}$ term models the 
exchange-repulsion energy, dominant at short internuclear distances, while the $R^{-6}$ term models the attractive vdW interaction, dominant at large internuclear distances. Whereas the attractive term is physically based, the 
repulsive term has no theoretical justification (i.e., chosen for computational efficiency). Note that the exchange-repulsion energy should decay almost exponentially with the internuclear distance. Nevertheless, due to its 
computational simplicity, the LJ potential is widely used in computer simulations even though more accurate potentials exist. 

However, the $R^{-6}$ dependence of the vdW interaction may not be applied to macroscopic systems like colloids and biological membranes. In these systems, the vdW interaction between two objects immersed in a medium 
is strongly influenced by the dielectric properties of the objects and the medium. Accordingly, the resulting vdW interaction can be very different from the conventional $R^{-6}$ expression \cite{n1,n2,n3}, and can be completely 
repulsive under certain conditions \cite{n1,n7,n4}. Several fascinating phenomena have been discovered in these non-$R^{-6}$ macroscopic vdW systems \cite{n4,n5,n6,n7}. 

Is it possible to create non-$R^{-6}$ vdW interactions between rare-gas atoms in vacuum? Conceptually, the types of interparticle interactions (nuclear-electron, electron-electron, and nuclear-nuclear interactions), traditionally 
given by the Coulomb interactions, should play a fundamental role in determining the properties of atoms and molecules. Hence, we expect that non-$R^{-6}$ vdW interactions can appear by tuning the effective interparticle 
interactions of rare-gas atoms in vacuum. As a proof of concept, in this work, we address how the nature of vdW interactions in rare-gas dimers (i.e., the simplest vdW systems) changes with varying interparticle interactions, 
using the highly accurate coupled-cluster theory associated with those interparticle interactions. The rest of this paper is organized as follows. In Section II, we describe our model systems and computational details. We compare 
the results obtained from the coupled-cluster theory with those obtained from different computational methods, and give our comments on the connection between this study and a popular scheme in density functional theory in 
Section III. Our conclusions are given in Section IV.

\section{Model Systems and Computational Details} 

For a system consisting of $M$ nuclei and $N$ electrons in the Born-Oppenheimer approximation (as the nuclei are much heavier than the electrons), the electronic Hamiltonian \cite{Jensen} 
\begin{equation} 
\begin{split} 
H_e = 
& - \frac{\hbar^2}{2 m_e} \sum_{i=1}^{N} {\bf \nabla}^2_i - \frac{e^2}{4 \pi \epsilon_0} \sum_{i=1}^{N} \sum_{A=1}^{M} Z_A f(r_{iA}) \\ 
& + \frac{e^2}{4 \pi \epsilon_0} \sum_{i=1}^{N} \sum_{j>i}^{N} f(r_{ij}) 
\end{split} 
\label{eq:He} 
\end{equation} 
is the sum of the kinetic energy of electrons, the nuclear-electron attraction energy, and the electron-electron repulsion energy, respectively. Here $Z_A$ is the atomic number of nucleus $A$, $m_e$ is the mass of an electron, 
$-e$ is the charge of an electron, $r_{iA} = |{\bf r}_i - {\bf R}_A|$ is the distance between electron $i$ and nucleus $A$, $r_{ij} = |{\bf r}_i - {\bf r}_j|$ is the distance between electrons $i$ and $j$, and $f(r)$ is the interparticle 
interaction operator with $r$ being the interparticle distance. The electronic Schr\"odinger equation 
\begin{equation} 
H_e \Psi_e = E_e \Psi_e 
\label{eq:SE} 
\end{equation} 
is solved for the electronic energy $E_e$ and electronic wavefunction $\Psi_e$, which describes the motion of the electrons for fixed nuclear positions. The total energy 
\begin{equation} 
E_{total} = E_e + \frac{e^2}{4 \pi \epsilon_0} \sum_{A=1}^{M} \sum_{B>A}^{M} Z_A Z_B f(R_{AB}) 
\label{eq:energy} 
\end{equation} 
is obtained by adding the nuclear-nuclear repulsion energy to the electronic energy, where $R_{AB} = |{\bf R}_A - {\bf R}_B|$ is the distance between nuclei $A$ and $B$. One can obtain $E_{total}$ as a function of the nuclear 
positions, commonly known as the potential energy curve (or surface). 

\begin{figure}
\includegraphics[scale=0.9]{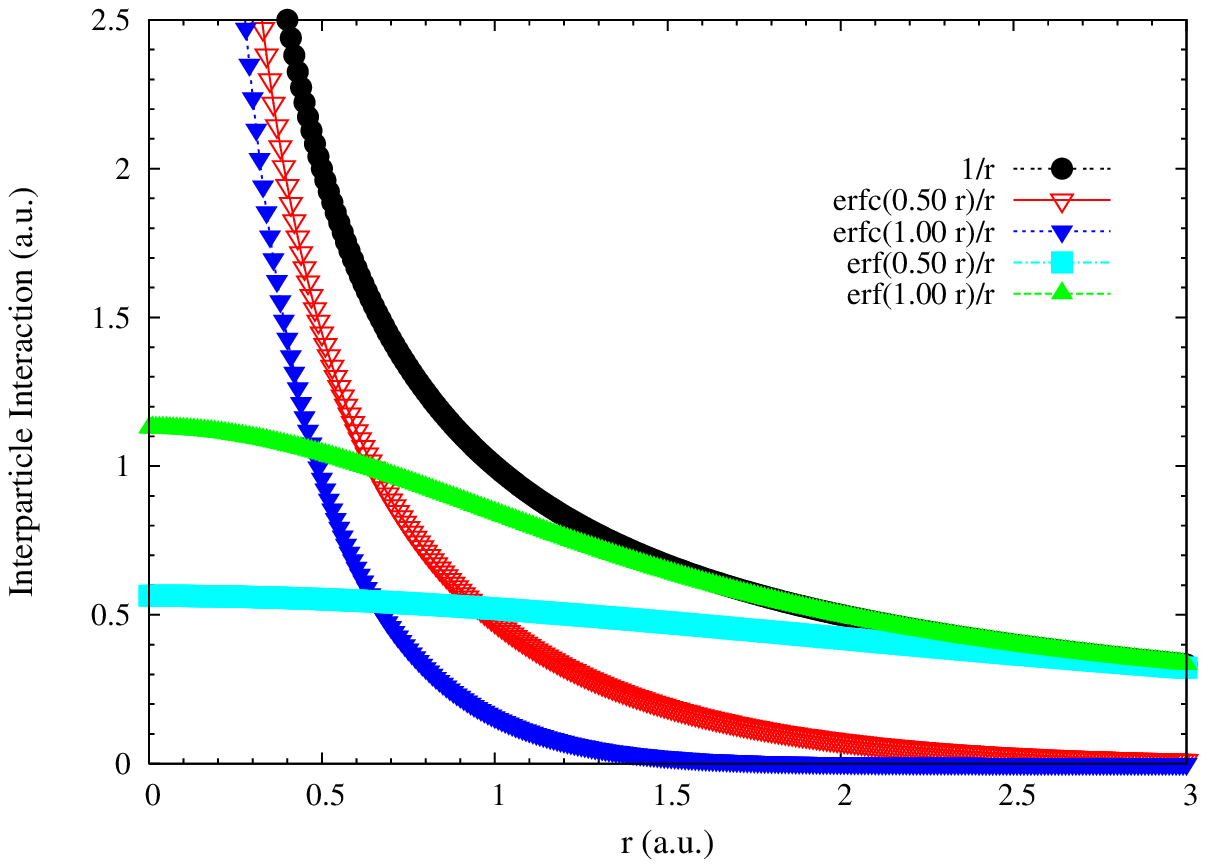} 
\caption{\label{fig:erfc_erf} 
Interparticle interaction as a function of interparticle distance (in atomic units).} 
\end{figure}

Traditionally, $f(r)$ is given by the Coulomb interaction $1/r$. However, in this work, we consider two types of $f(r)$: $\text{erf}(\omega r)/r$ and $\text{erfc}(\omega r)/r$, which are generated by splitting the Coulomb interaction 
into two components \cite{CASE,CASE_MP2}. The former (the erf interaction) retains the long-range Coulomb tail without the singularity at $r = 0$, while the latter (the erfc interaction) is a short-range interaction with a singularity 
at $r = 0$. Physically, $1/\omega$ specifies the distance beyond which $\text{erf}(\omega r)/r$ approaches $1/r$ and the distance beyond which $\text{erfc}(\omega r)/r$ becomes insignificant (see \Cref{fig:erfc_erf}). Similar to 
the Coulomb case \cite{Boys,Gauss}, the nuclear-attraction and two-electron repulsion integrals modified for the erf and erfc interaction operators can be evaluated analytically over Gaussian basis 
functions \cite{Gauss,erf_anal,erf_anal2}, facilitating an efficient evaluation of the integrals needed for solving Eq.\ (\ref{eq:SE}) and the equations associated with related approximate methods (see below). In principle, other 
types of $f(r)$ can also be adopted \cite{erf_anal,erfgau_anal,terf_anal,op,terf_JAP}. 

Similar to the Coulomb case, solving Eq.\ (\ref{eq:SE}) for a given $f(r)$ is, however, extremely difficult even for the ground-state energy and wavefunction of a very small system, due to the prohibitively expensive computational 
cost. Practically, one searches for approximate solutions to Eq.\ (\ref{eq:SE}), obtained from {\it ab initio} wavefunction methods \cite{Jensen,CASE,CASE_MP2,PC}, such as the Hartree-Fock (HF) theory, second-order 
M\o ller-Plesset perturbation theory (MP2), coupled-cluster theory with iterative singles and doubles (CCSD), and CCSD with a perturbative treatment of triple substitutions (CCSD(T)). Among them, the CCSD(T) method with a 
sufficiently large basis set is generally expected to provide highly accurate results for a variety of small- to medium-sized systems. 

Alternatively, Kohn-Sham density functional theory (KS-DFT) \cite{KS}, a popular method for the study of the ground-state properties of large systems, can also be employed. Similar to the Coulomb case, density functional 
approximations (DFAs), such as the local density approximation (LDA) and generalized-gradient approximations (GGAs), to the exchange-correlation (XC) energy functional for a given $f(r)$ are needed in the corresponding 
KS-DFT \cite{OD,RevYang}. Here, the LDA exchange energy functional for the erf interaction is obtained by subtracting the LDA exchange energy functional for the erfc interaction \cite{SR-LDAX} from the LDA exchange energy 
functional for the Coulomb interaction \cite{LDAX}, whereas the LDA correlation energy functional for the erfc interaction is obtained by subtracting the LDA correlation energy functional for the erf interaction \cite{LR-LDAC} from the 
LDA correlation energy functional for the Coulomb interaction \cite{LDAC}. In addition, as the Perdew-Burke-Ernzerhof (PBE) XC energy functional (i.e., a popular GGA) for the Coulomb interaction \cite{PBE} and its variant for the 
erfc interaction \cite{SR-PBE} are both available, their difference gives the PBE XC energy functional for the erf interaction. 

To illustrate how the nature of vdW interactions in rare-gas dimers changes with varying interparticle interactions, we calculate the potential energy curves of the He-He dimer associated with the interparticle interactions 
$\text{erf}(\omega r)/r$ ($\omega$ = $\infty$, 10.00, 2.00, 1.70, 1.40, and 1.10 bohr$^{-1}$) and $\text{erfc}(\omega r)/r$ ($\omega$ = 0.00, 0.10, 0.20, 0.25, 0.30, and 0.40 bohr$^{-1}$), using the corresponding CCSD(T), CCSD, 
MP2, HF, and KS-DFT employing the PBE and LDA XC energy functionals for the associated interactions \cite{Jensen,CASE,CASE_MP2,PC}. 

All calculations are performed with a development version of \textsf{Q-Chem 4.0} \cite{Q-Chem}. Results are computed using a large aug-cc-pVQZ basis set \cite{AQZ} with a high-quality EML(250,590) grid, consisting of 
250 Euler-Maclaurin radial grid points \cite{EM} and 590 Lebedev angular grid points \cite{L}. The counterpoise correction \cite{CP} is employed to reduce basis set superposition error (BSSE).

\section{Results and Discussion} 

\begin{figure}
\includegraphics[scale=1.0]{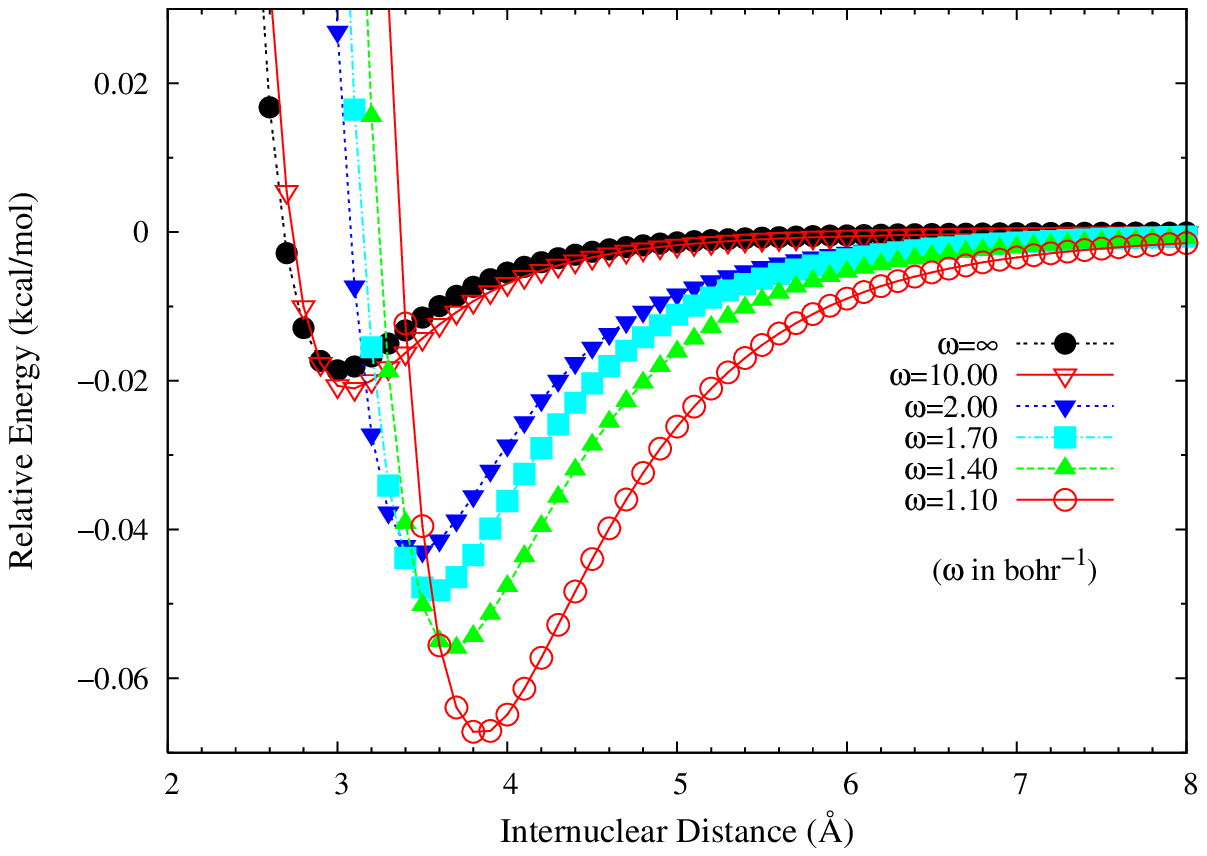}
\caption{\label{fig:He2erf} 
Potential energy curves of the He-He dimer associated with the long-range interparticle interactions $\text{erf}(\omega r)/r$, calculated using the corresponding CCSD(T). 
The $\omega=\infty$ case is equivalent to the Coulomb interaction $1/r$.} 
\end{figure}

The potential energy curves of the He-He dimer associated with the long-range interparticle interactions $\text{erf}(\omega r)/r$, calculated using the corresponding CCSD(T), are presented in \Cref{fig:He2erf}. Similar to the 
Coulomb case (i.e., the $\omega=\infty$ case of the erf interaction), all the potential energy curves resemble the LJ potentials. For a smaller $\omega$, the strength of the erf interaction is weaker. Consequently, the electrons are 
more loosely bound to the nucleus, and the atoms are more polarizable, yielding larger values of $\sigma$ and $\epsilon$, respectively (see Eq.\ (\ref{eq:LJ})) \cite{Jensen}. Owing to the long-range nature of the erf interaction, the 
attractive vdW interaction is shown to have the $[\text{erf}(\omega R)]^2 R^{-6}$ asymptote (essentially retaining the $R^{-6}$ asymptote of conventional vdW interactions) at sufficiently large internuclear distances $R$, based on 
a second-order perturbation theory (see the Appendix). 

\begin{figure} 
\subfigure 
{\includegraphics[scale=0.6,trim = 0mm 0mm 0mm 0mm, clip=true, clip=true]{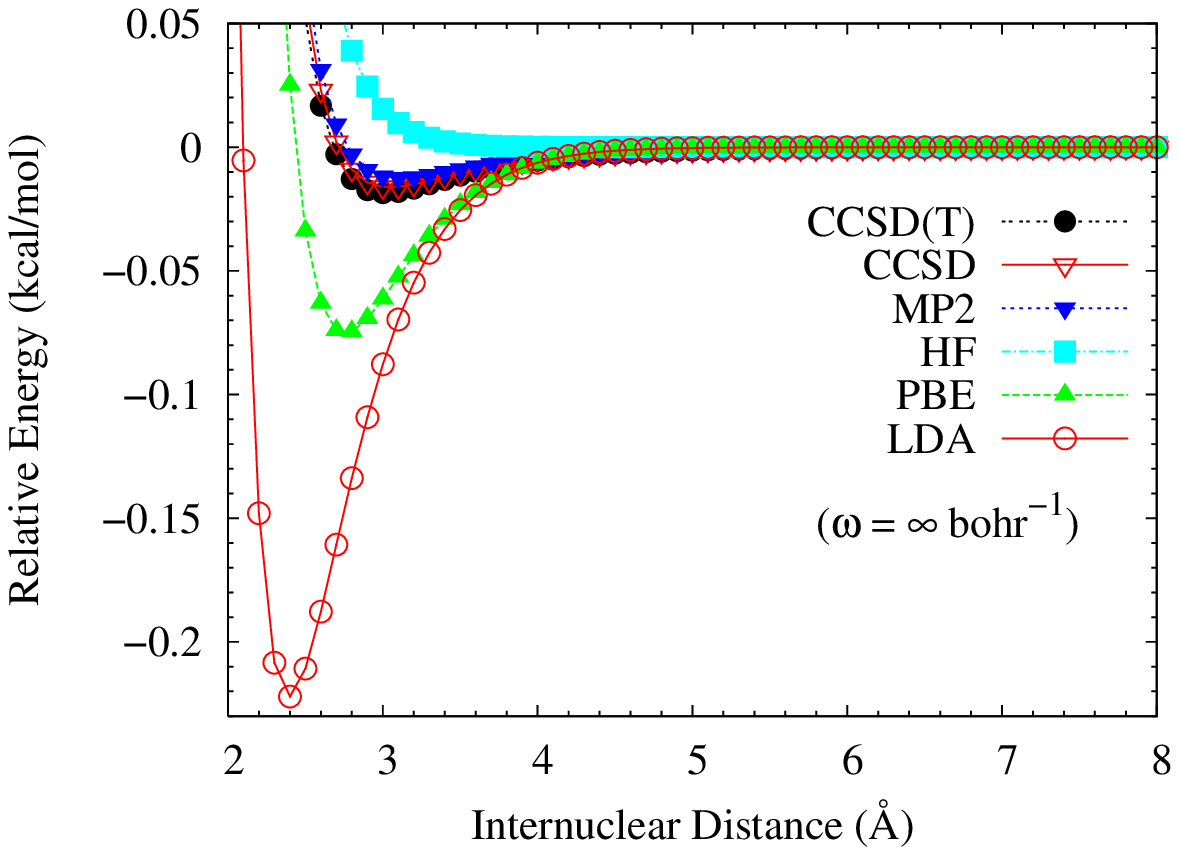}} 
\subfigure 
{\includegraphics[scale=0.6,trim = 0mm 0mm 0mm 0mm, clip=true, clip=true]{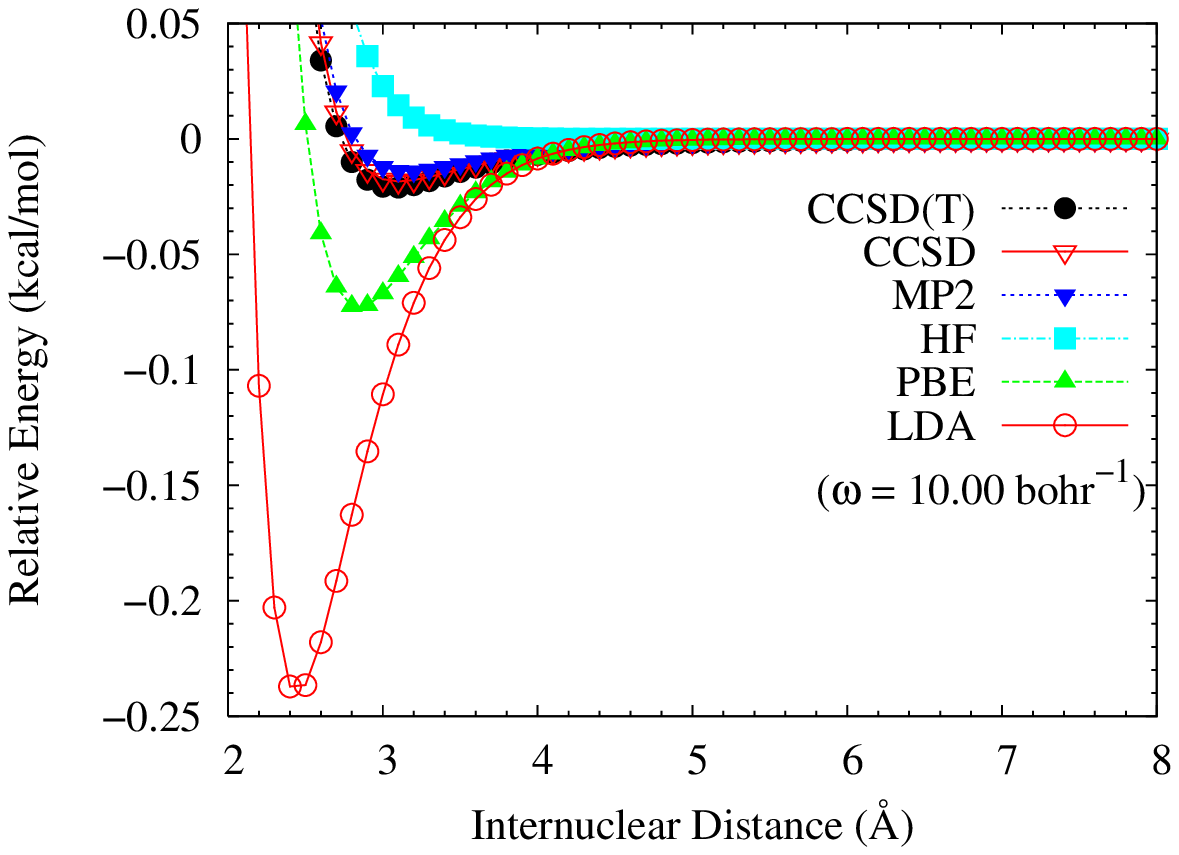}} 
\subfigure 
{\includegraphics[scale=0.6,trim = 0mm 0mm 0mm 0mm, clip=true, clip=true]{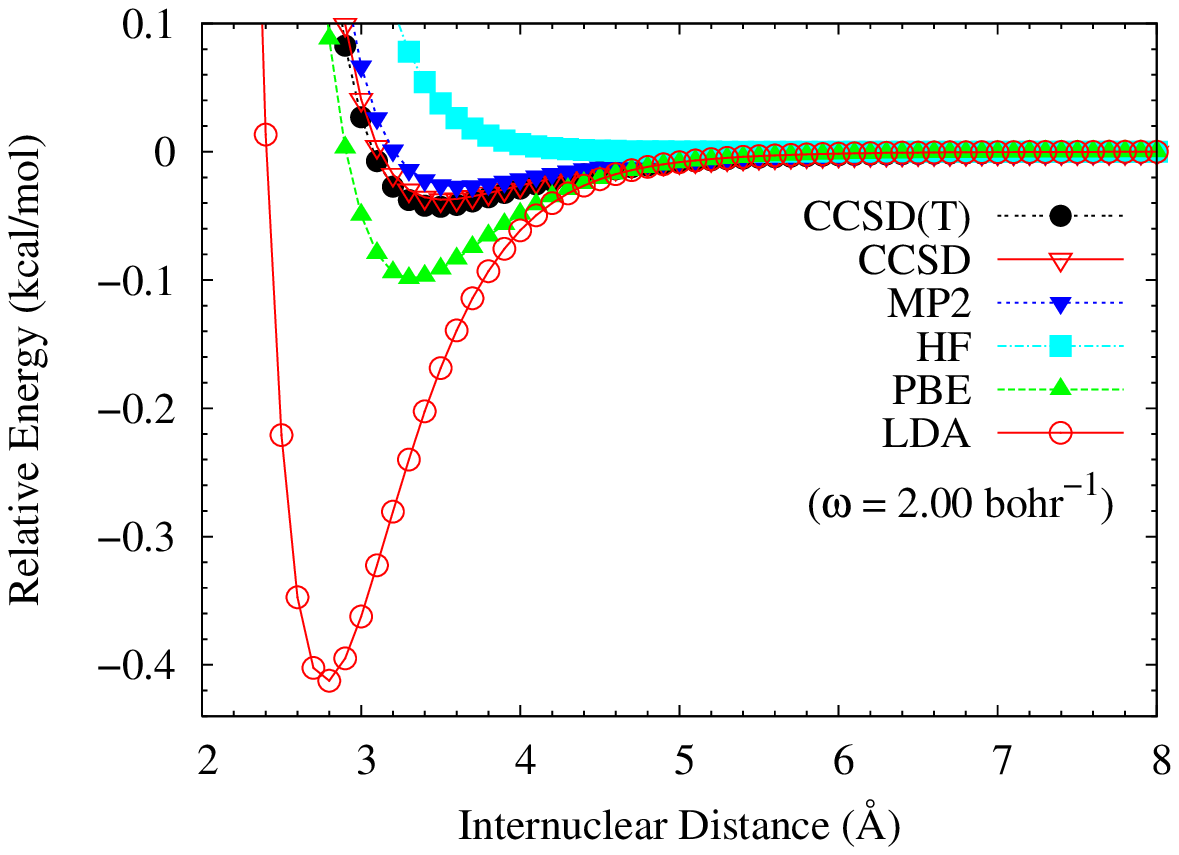}} 
\subfigure 
{\includegraphics[scale=0.6,trim = 0mm 0mm 0mm 0mm, clip=true, clip=true]{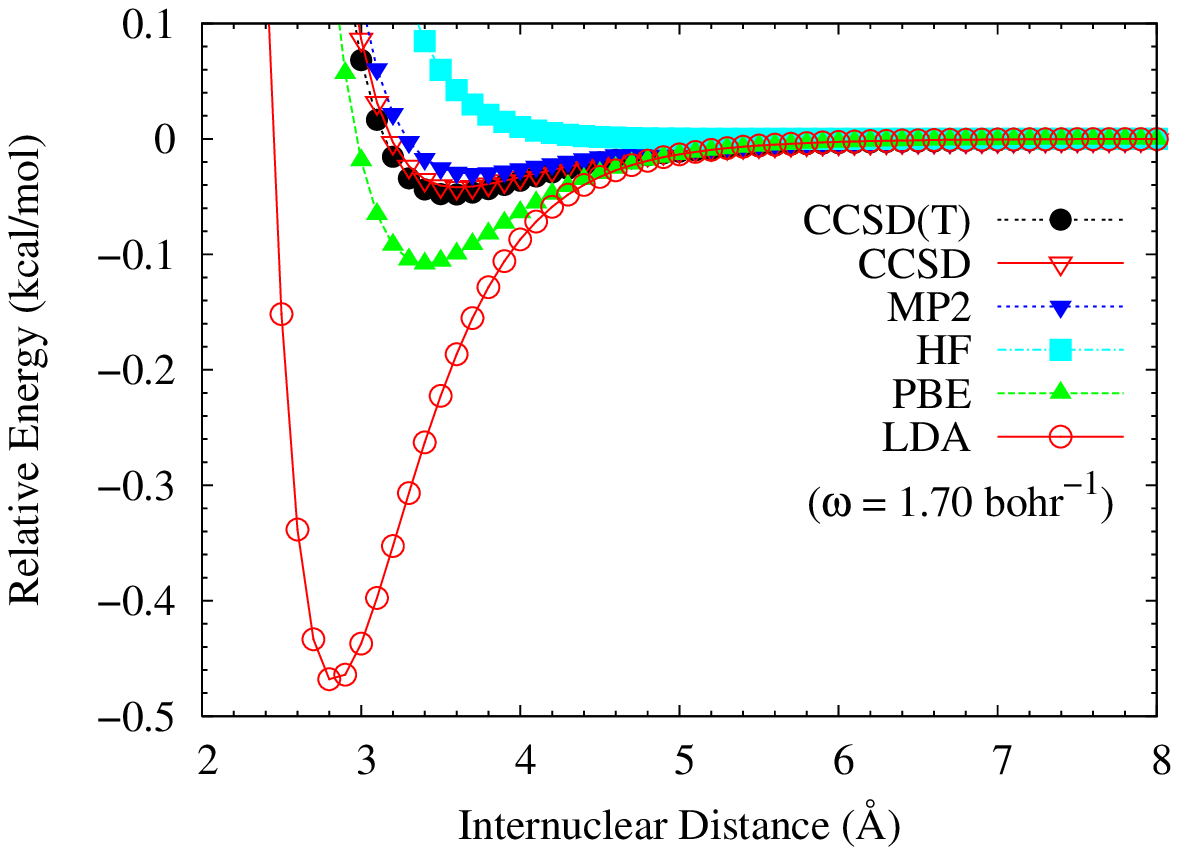}} 
\subfigure 
{\includegraphics[scale=0.6,trim = 0mm 0mm 0mm 0mm, clip=true, clip=true]{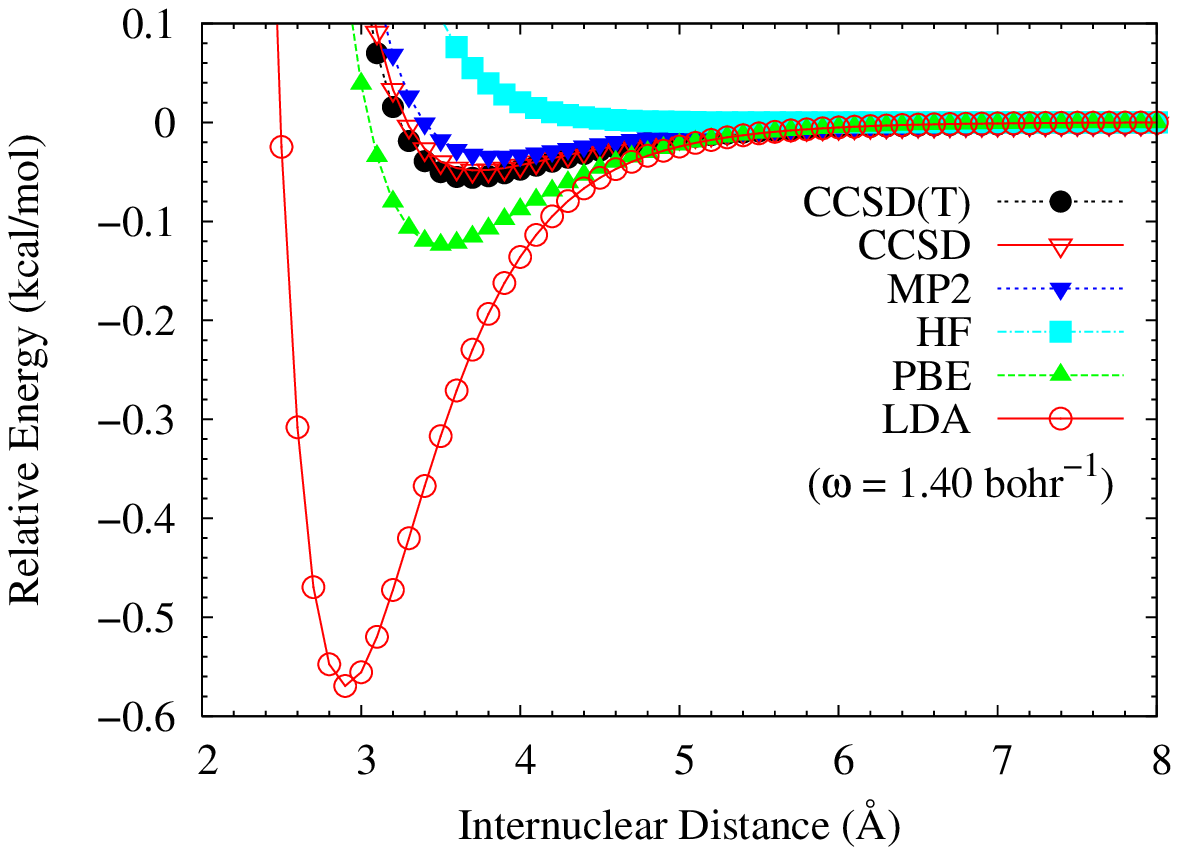}} 
\subfigure 
{\includegraphics[scale=0.6,trim = 0mm 0mm 0mm 0mm, clip=true, clip=true]{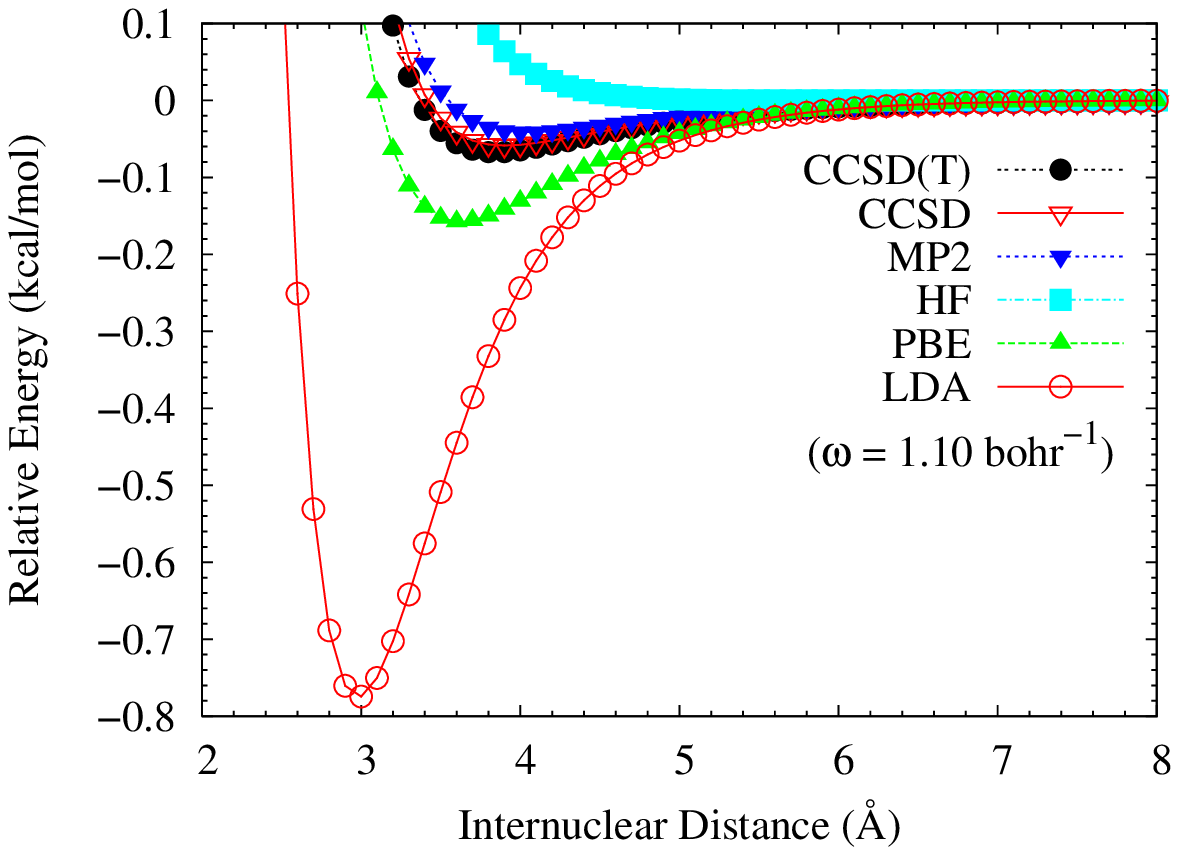}} 
\caption{\label{fig:He2erf2} 
Potential energy curves of the He-He dimer associated with the long-range interparticle interactions $\text{erf}(\omega r)/r$, calculated using the corresponding CCSD(T), CCSD, MP2, HF, PBE, and LDA. 
The $\omega=\infty$ case is equivalent to the Coulomb interaction $1/r$.} 
\end{figure}

In comparison with the highly accurate CCSD(T) results, the He-He potential energy curves associated with the erf interactions, calculated using the corresponding CCSD, MP2, HF, PBE, and LDA are presented in \Cref{fig:He2erf2}. 
As shown, CCSD performs similarly to CCSD(T), and slightly outperforms MP2. Besides, CCSD(T), CCSD, and MP2 exhibit the correct $R^{-6}$ vdW asymptotes. By contrast, due to the lack of electron correlation, HF completely 
fails to describe the attractive vdW interactions, yielding purely repulsive potential energy curves for all the $\omega$ values studied. Within the framework of KS-DFT, PBE consistently outperforms LDA. However, in view of the 
large errors associated with the vdW minima and the incorrect vdW asymptotes (decaying much faster than $R^{-6}$) \cite{vdWDF,vdWasymp}, LDA, PBE, and possibly other semilocal density functionals \cite{semilocal} cannot 
accurately describe long-range vdW interactions \cite{Dobson}, wherein a fully nonlocal XC energy functional should be essential \cite{OD,RevYang,vdWDF}. 

\begin{figure} 
\includegraphics[scale=1.0]{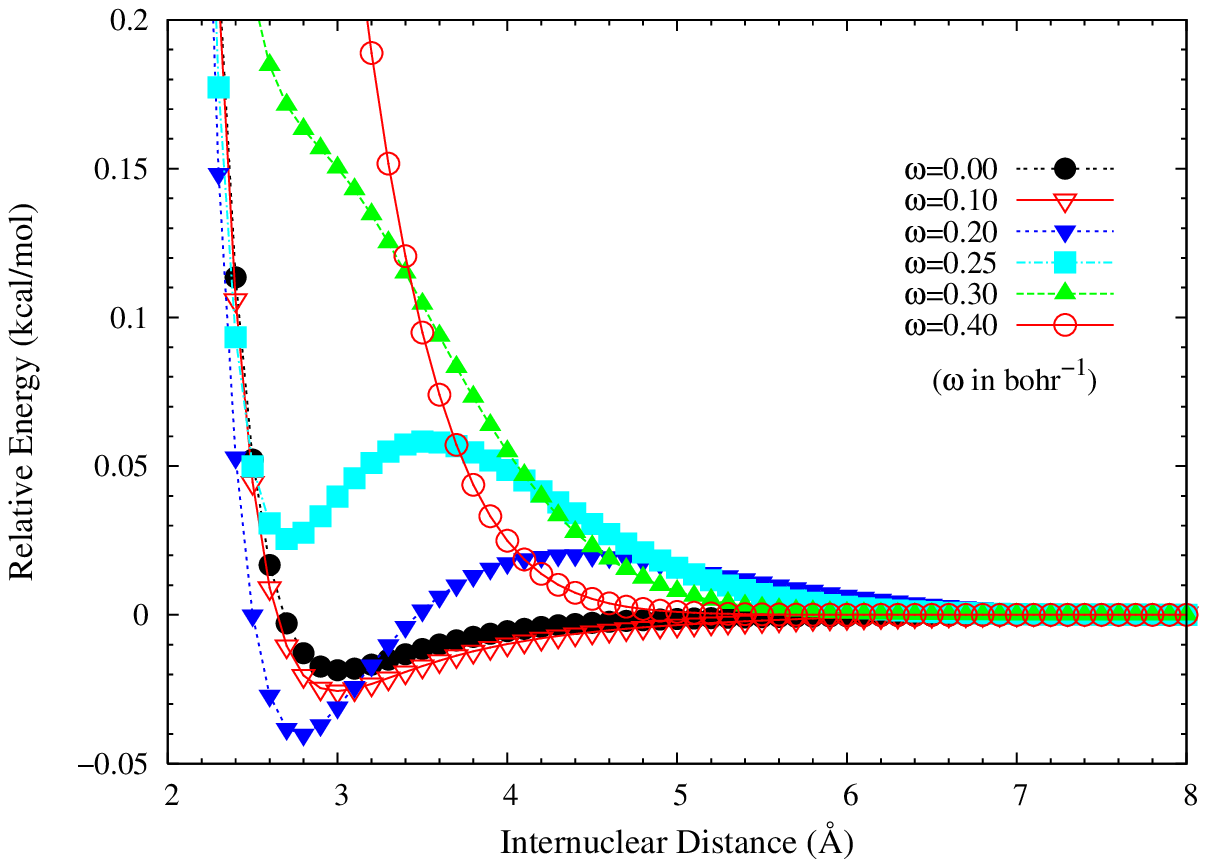} 
\caption{\label{fig:He2erfc} 
Potential energy curves of the He-He dimer associated with the short-range interparticle interactions $\text{erfc}(\omega r)/r$, calculated using the corresponding CCSD(T). 
The $\omega=0$ case is equivalent to the Coulomb interaction $1/r$.} 
\end{figure}

On the other hand, the potential energy curves of the He-He dimer associated with the short-range interparticle interactions $\text{erfc}(\omega r)/r$, calculated using the corresponding CCSD(T), are shown in \Cref{fig:He2erfc}. 
In contrast to the Coulomb case (i.e., the $\omega=0$ case of the erfc interaction), the potential energy curves show strong $\omega$-dependence. It resembles the LJ potential only for a vanishingly small $\omega$, displays a 
metastable state for an intermediate $\omega$ (around 0.25 bohr$^{-1}$ or smaller), and becomes purely repulsive for a $\omega$ larger than 0.30 bohr$^{-1}$ (see the Appendix). 

\begin{figure} 
\subfigure 
{\includegraphics[scale=0.6,trim = 0mm 0mm 0mm 0mm, clip=true, clip=true]{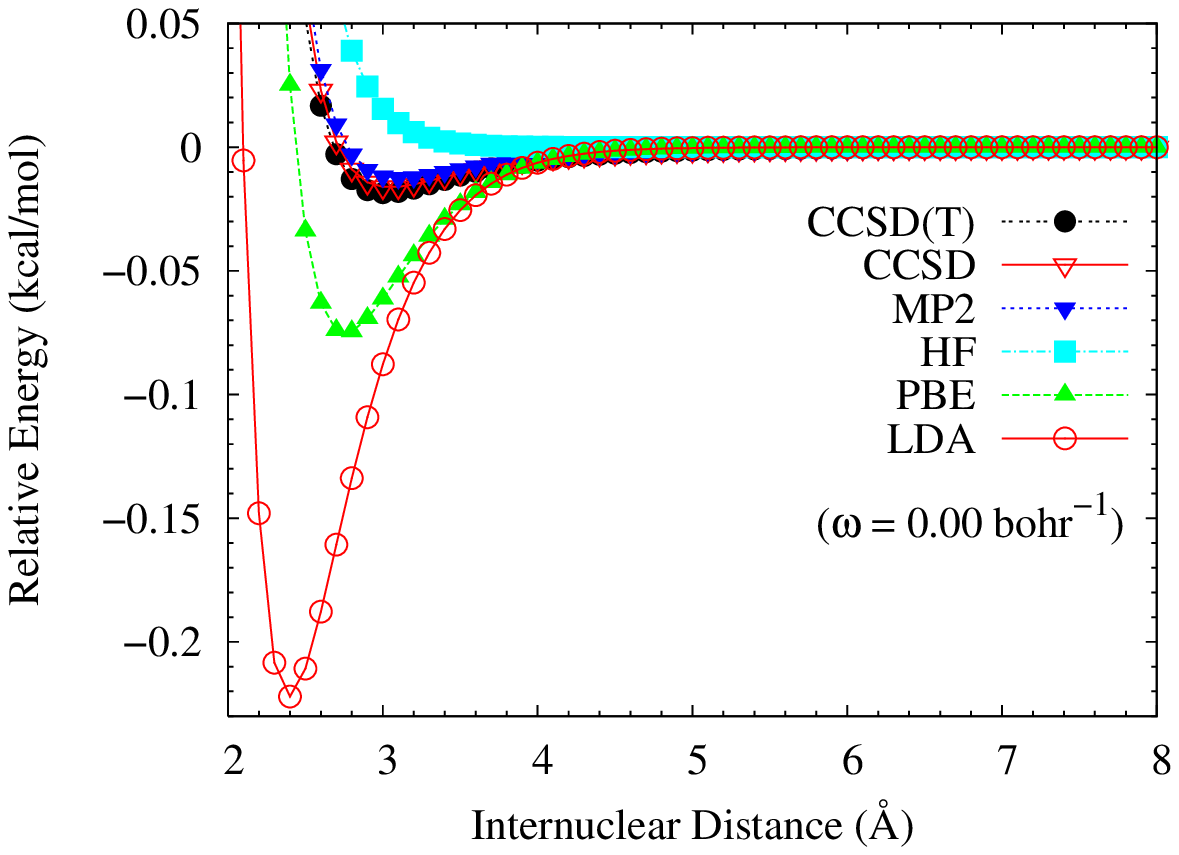}} 
\subfigure 
{\includegraphics[scale=0.6,trim = 0mm 0mm 0mm 0mm, clip=true, clip=true]{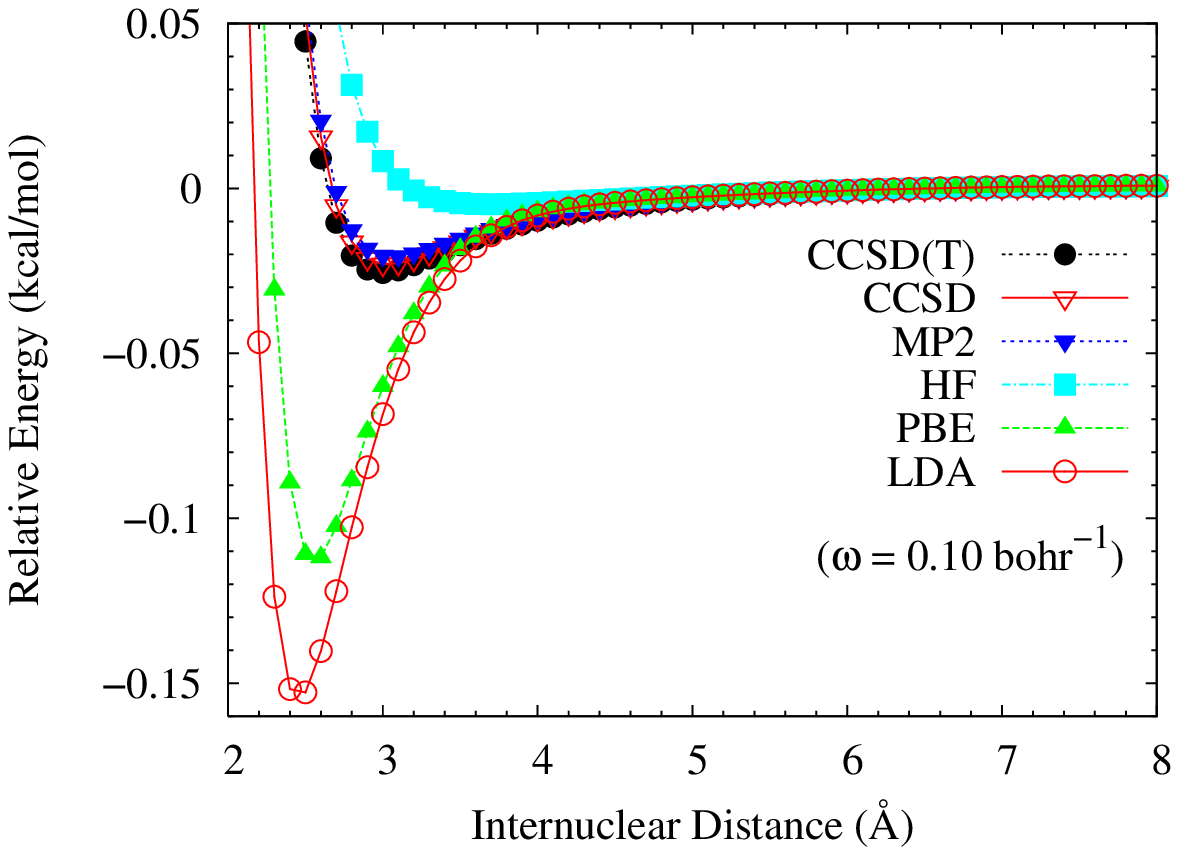}} 
\subfigure 
{\includegraphics[scale=0.6,trim = 0mm 0mm 0mm 0mm, clip=true, clip=true]{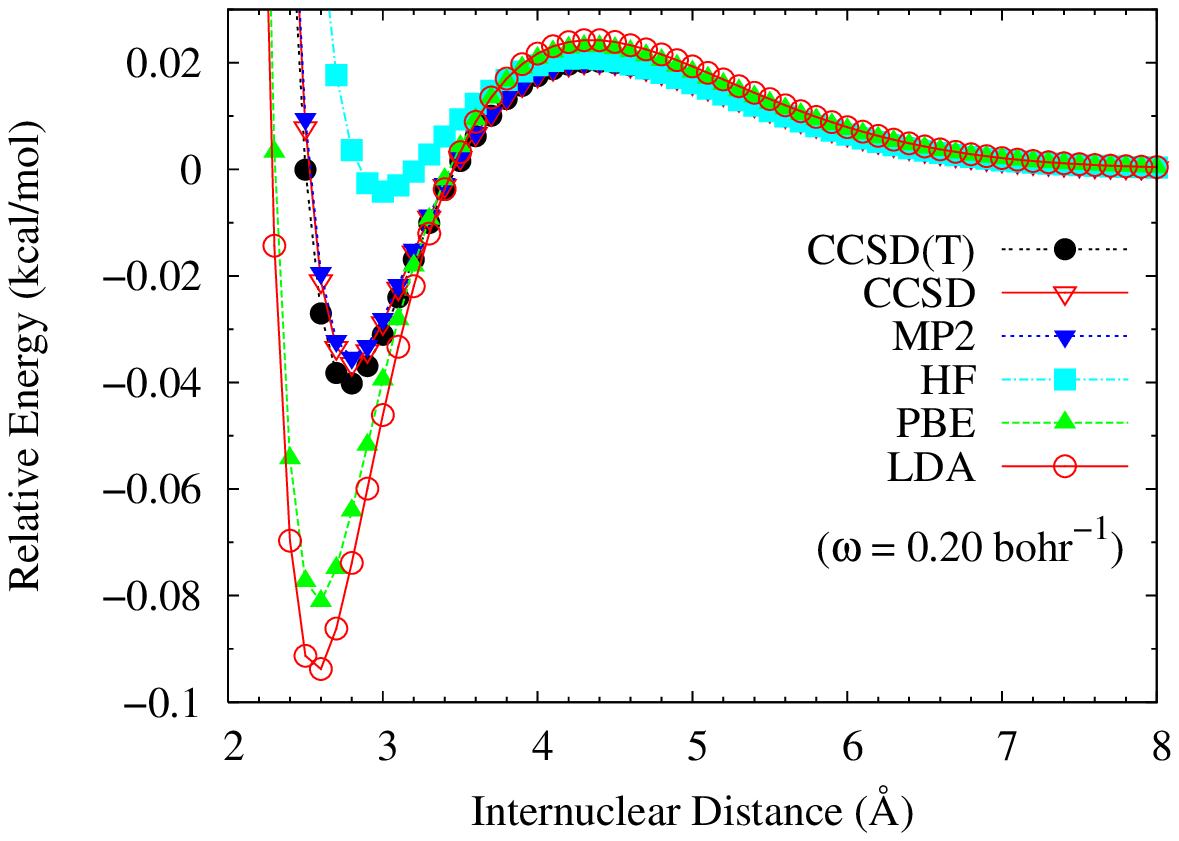}} 
\subfigure 
{\includegraphics[scale=0.6,trim = 0mm 0mm 0mm 0mm, clip=true, clip=true]{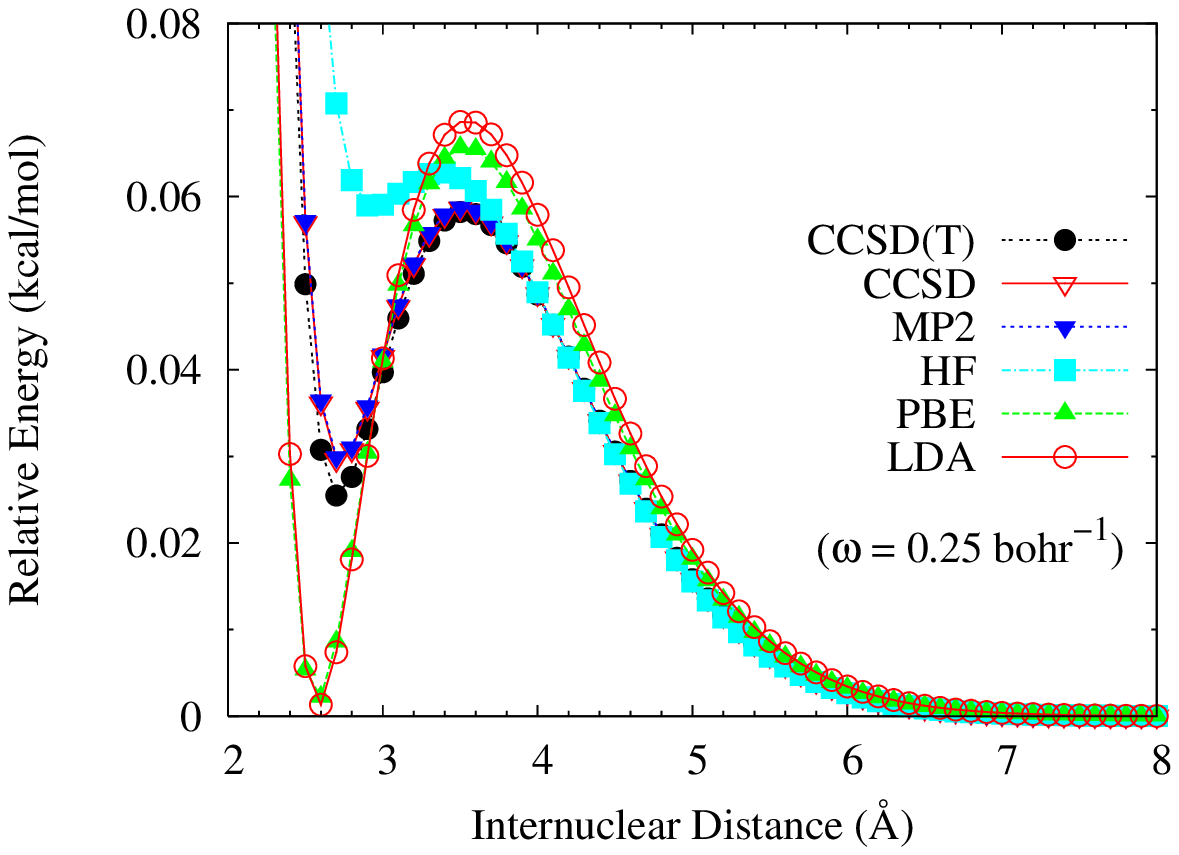}}  
\subfigure 
{\includegraphics[scale=0.6,trim = 0mm 0mm 0mm 0mm, clip=true, clip=true]{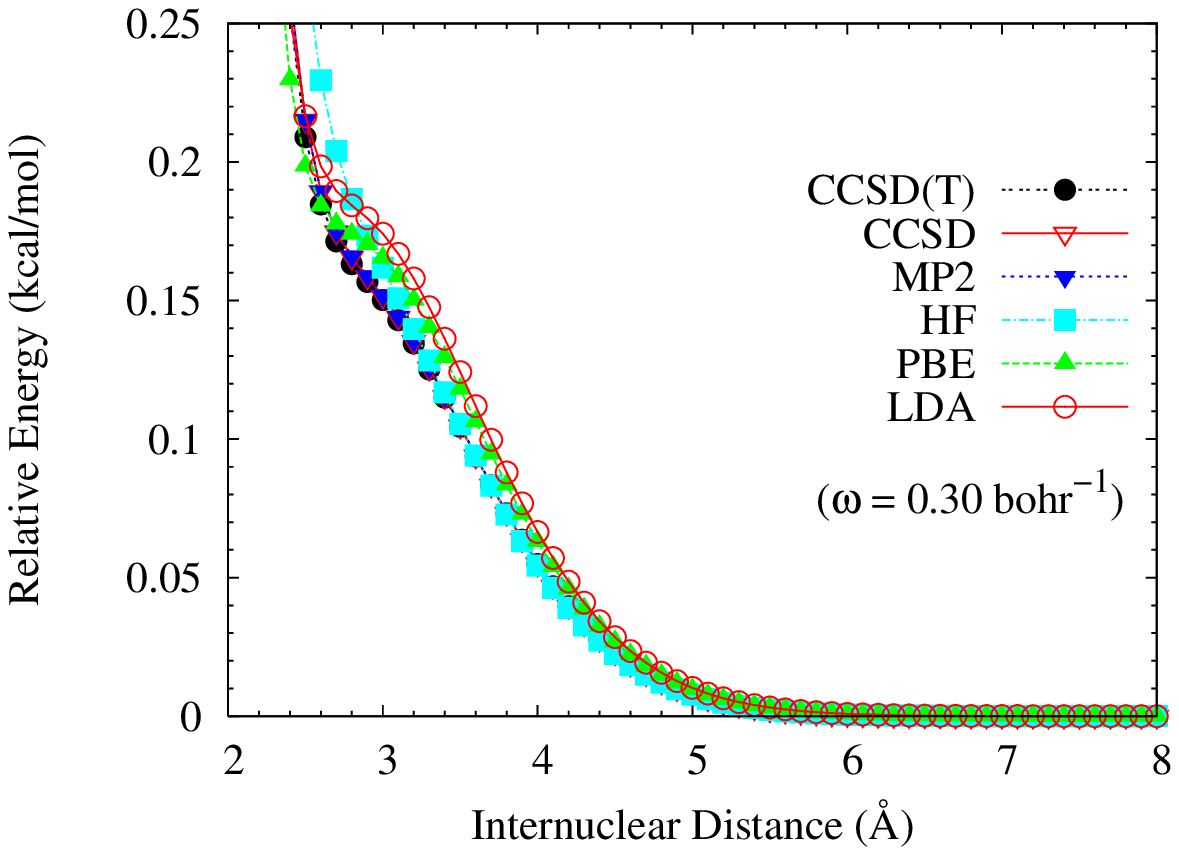}} 
\subfigure 
{\includegraphics[scale=0.6,trim = 0mm 0mm 0mm 0mm, clip=true, clip=true]{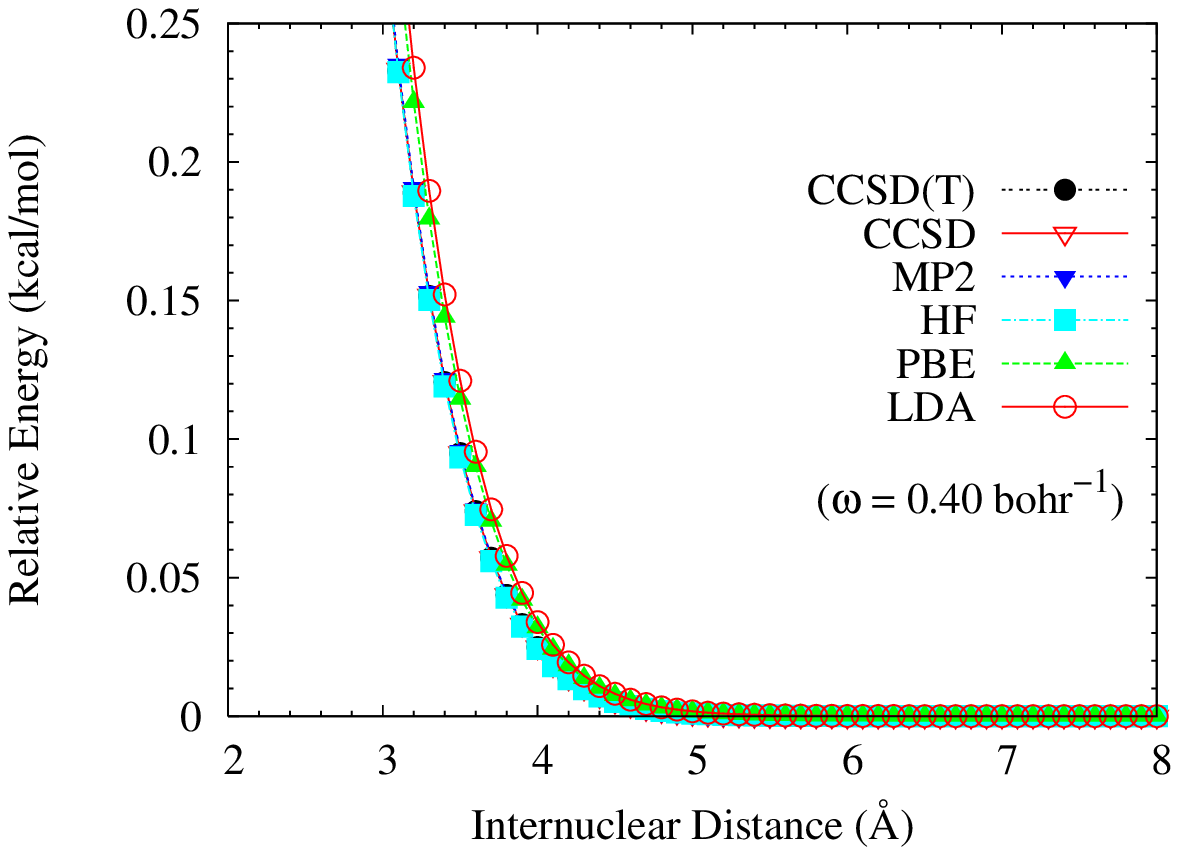}} 
\caption{\label{fig:He2erfc2} 
Potential energy curves of the He-He dimer associated with the short-range interparticle interactions $\text{erfc}(\omega r)/r$, calculated using the corresponding CCSD(T), CCSD, MP2, HF, PBE, and LDA. 
The $\omega=0$ case is equivalent to the Coulomb interaction $1/r$.} 
\end{figure}

For comparison, the He-He potential energy curves associated with the erfc interactions, calculated using the corresponding CCSD, MP2, HF, PBE, and LDA are shown in \Cref{fig:He2erfc2}. With the increase of $\omega$, the 
potential energy curves obtained from all the computational methods become very similar. As would be expected on physical grounds, semilocal density functionals can be surprisingly accurate for short-range XC 
effects \cite{semilocal}. PBE is shown to consistently perform better than LDA. Besides, due to the dominance of exchange-repulsion energy for a sufficiently large $\omega$, even the HF theory can be reliably accurate. 

Similar to the Coulomb case, the overall trends of LDA and PBE are opposite to those of HF and MP2, implying that a combination of the HF exchange, MP2 correlation, and DFAs (e.g., LDA or GGAs) in KS-DFT (i.e., hybrid 
DFT \cite{hybrid} or double-hybrid DFT \cite{B2PLYP}) may achieve a more favorable balance between cost and performance than CCSD(T) for the vdW interactions in large rare-gas dimers under the erf and erfc interactions. 

\begin{figure} 
\includegraphics[scale=0.9]{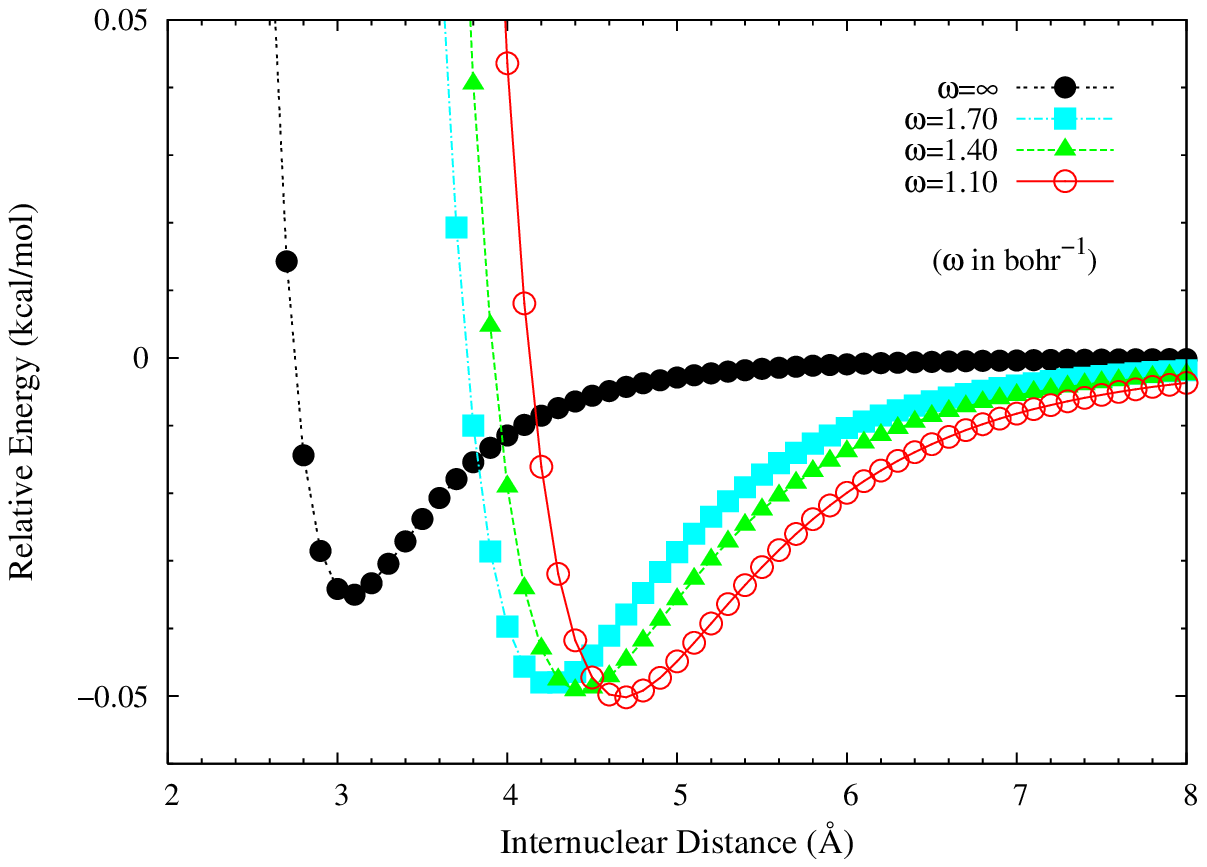} 
\caption{\label{fig:He-Neerfccsdt} 
Potential energy curves of the He-Ne dimer associated with the long-range interparticle interactions $\text{erf}(\omega r)/r$, calculated using the corresponding CCSD(T). 
The $\omega=\infty$ case is equivalent to the Coulomb interaction $1/r$.} 
\end{figure}

\begin{figure} 
\includegraphics[scale=0.9]{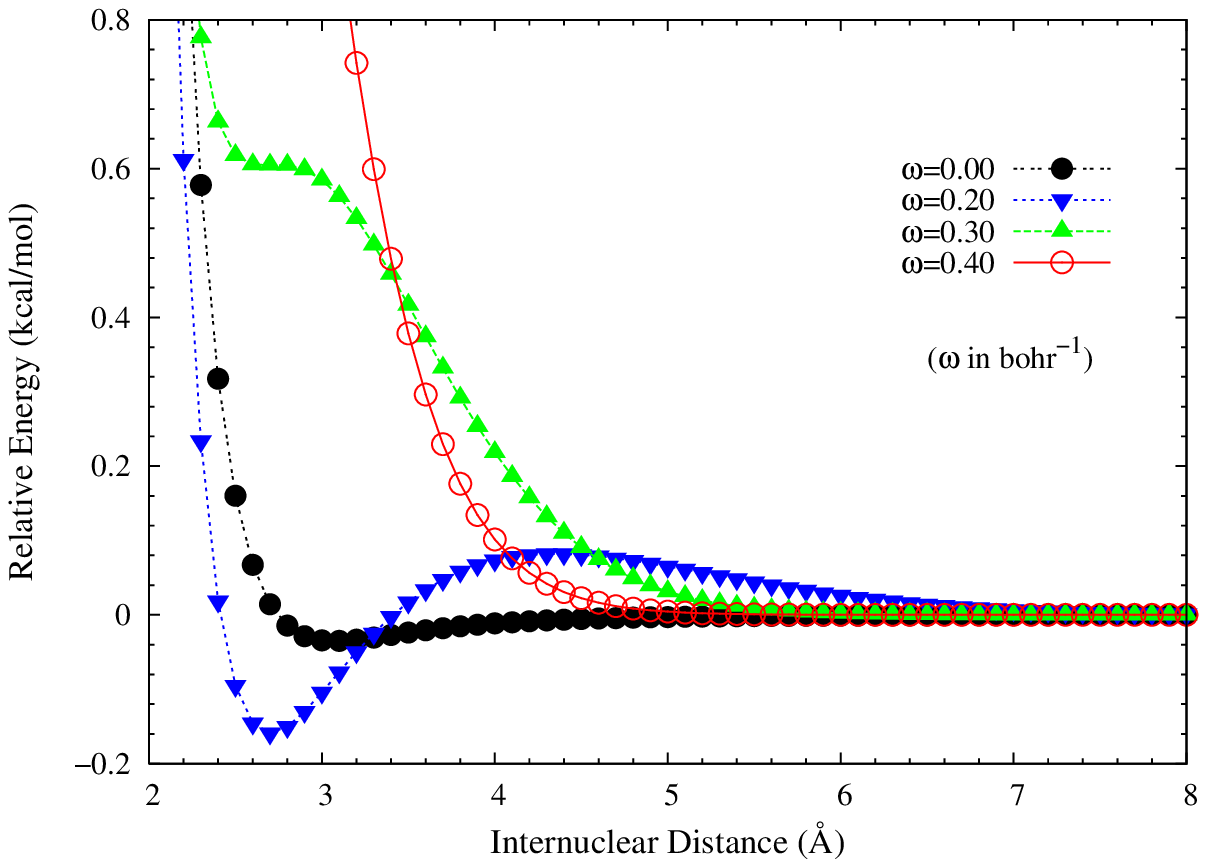} 
\caption{\label{fig:He-Neerfcccsdt} 
Potential energy curves of the He-Ne dimer associated with the short-range interparticle interactions $\text{erfc}(\omega r)/r$, calculated using the corresponding CCSD(T). 
The $\omega=0$ case is equivalent to the Coulomb interaction $1/r$.} 
\end{figure}

\begin{figure} 
\includegraphics[scale=0.9]{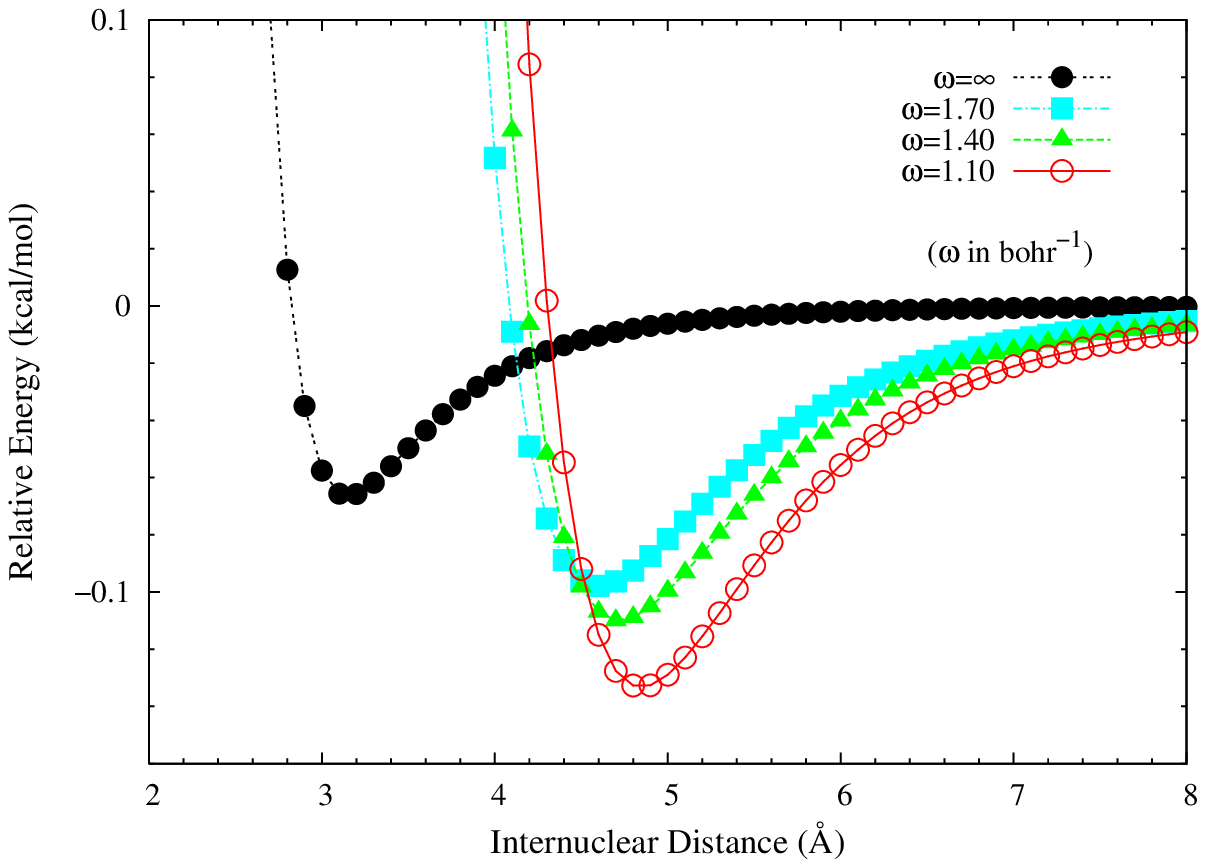} 
\caption{\label{fig:Ne-Neerfccsdt} 
Potential energy curves of the Ne-Ne dimer associated with the long-range interparticle interactions $\text{erf}(\omega r)/r$, calculated using the corresponding CCSD(T). 
The $\omega=\infty$ case is equivalent to the Coulomb interaction $1/r$.} 
\end{figure}

\begin{figure} 
\includegraphics[scale=0.9]{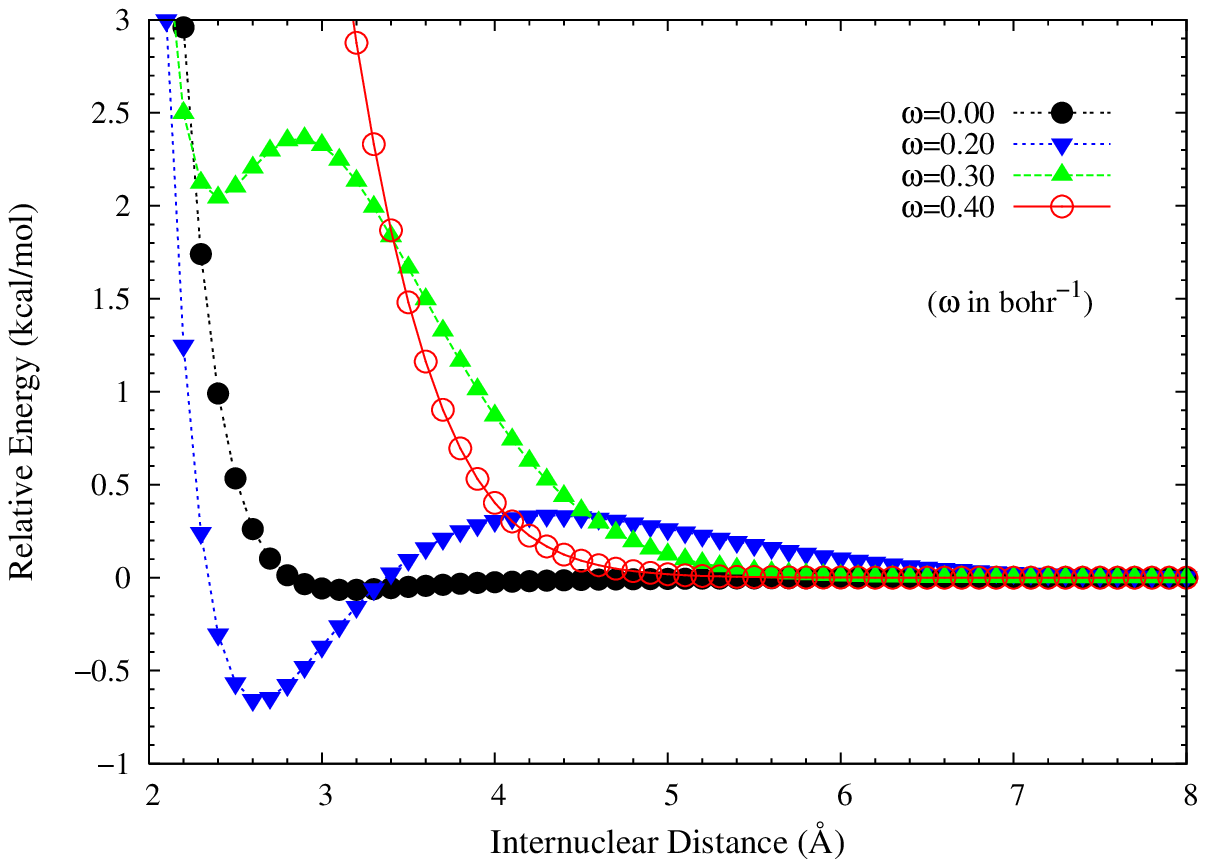} 
\caption{\label{fig:Ne-Neerfcccsdt} 
Potential energy curves of the Ne-Ne dimer associated with the short-range interparticle interactions $\text{erfc}(\omega r)/r$, calculated using the corresponding CCSD(T). 
The $\omega=0$ case is equivalent to the Coulomb interaction $1/r$.} 
\end{figure}

In addition, we calculate the potential energy curves of the He-Ne and Ne-Ne dimers associated with the erf and erfc interactions, using the corresponding CCSD(T), as shown in 
\Cref{fig:He-Neerfccsdt,fig:He-Neerfcccsdt,fig:Ne-Neerfccsdt,fig:Ne-Neerfcccsdt}. For the Coulomb case, the values of $\sigma$ for the He-Ne and Ne-Ne dimers are larger than that for the He-He dimer. Nevertheless, similar 
trends are also found for the potential energy curves of the He-Ne, Ne-Ne, and possibly other rare-gas dimers. 

\begin{figure} 
\includegraphics[scale=1.0]{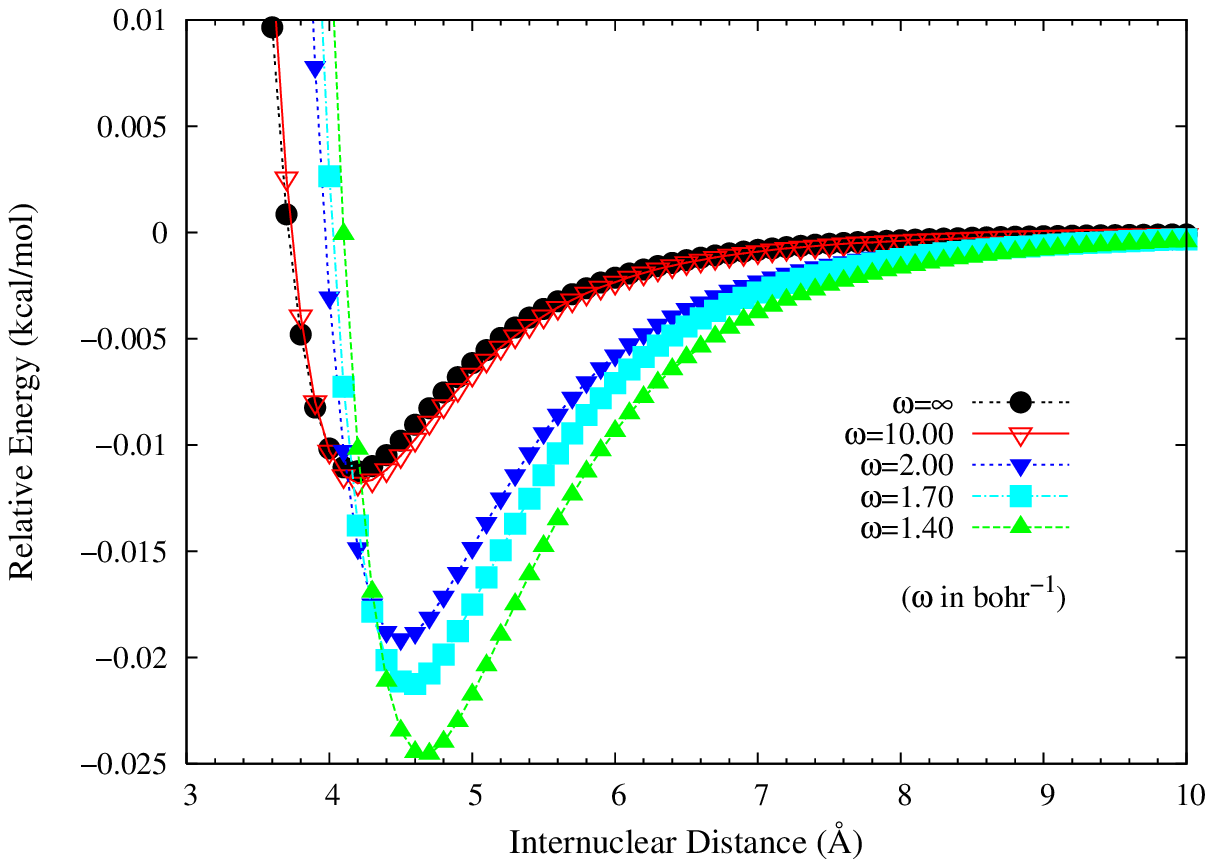} 
\caption{\label{fig:tH2erf} 
Potential energy curves for the lowest triplet states of H$_2$ associated with the long-range interparticle interactions $\text{erf}(\omega r)/r$, calculated using the corresponding CCSD. 
The $\omega=\infty$ case is equivalent to the Coulomb interaction $1/r$.} 
\end{figure}

\begin{figure} 
\includegraphics[scale=1.0]{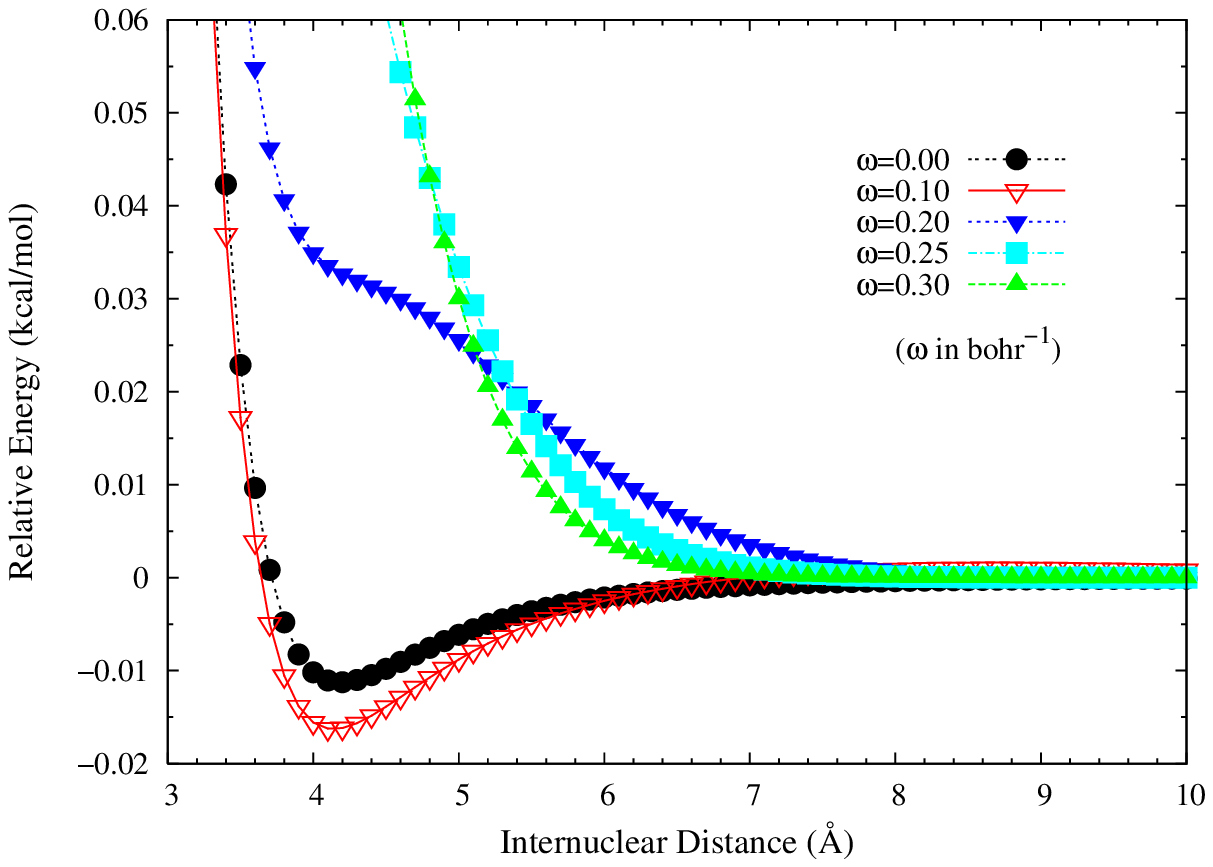} 
\caption{\label{fig:tH2erfc} 
Potential energy curves for the lowest triplet states of H$_2$ associated with the short-range interparticle interactions $\text{erfc}(\omega r)/r$, calculated using the corresponding CCSD. 
The $\omega=0$ case is equivalent to the Coulomb interaction $1/r$.} 
\end{figure}

To test the {\it transferability} of the above observed trends for other vdW systems, we calculate the potential energy curves for the lowest triplet states of H$_2$ (a simple vdW system) \cite{tH2} associated with the erf and erfc 
interactions, using the corresponding CCSD (i.e., an exact theory for any two-electron system). As shown in \Cref{fig:tH2erf,fig:tH2erfc}, the major features of the potential energy curves remain very similar to those found for 
rare-gas dimers. 

Here we comment on the connection between this study and long-range corrected (LC) hybrid functionals for systems with Coulomb interactions \cite{LC-DFT,LCHirao,CAM-B3LYP,LC-wPBE,BNL,wB97X,wB97X-D,Herbert,
wB97X-2,M11,wM05-D,LC-D3}. These functionals model the short-range interaction (e.g., the erfc interaction) by a DFA in KS-DFT and the complementary long-range interaction (e.g., the erf interaction) by HF exchange or a 
fully nonlocal (i.e., orbital-dependent) XC energy component from {\it ab initio} wavefunction methods. In \Cref{fig:He2erf2,fig:He2erfc2}, compared to the highly accurate CCSD(T) results, LDA and PBE perform reasonably well 
for sufficiently short-range interparticle interactions, whereas they perform poorly for long-range interparticle interactions. Accordingly, our findings are also in support of the key feature of the LC hybrid functionals for systems 
with Coulomb interactions, which have recently been found to provide supreme performance for a very wide range of applications \cite{LCAC,EB}, especially for problems related to the asymptote of the XC 
potential \cite{exactip,asymp,LB94,sum_vxc2,sum_vxc3,Staroverov,AC1,LFA}, self-interaction errors \cite{SIC,SIE}, fundamental 
gaps \cite{DD1,DD2,DD2a,DD4a,DDcorr,DD6,DD7,errsinfuncs,DD8,DDHirao,GKS2,Correction2,Correction3,DDChai}, and charge-transfer excitations \cite{Dreuw,Tozer,Gritsenko,Dreuw2,Dreuw3,fxcd,Peng}. Besides, empirical 
atom-atom dispersion potentials \cite{DFT-D1,DFT-D2,DFT-D3,wB97X-D,wM05-D,LC-D3} or MP2 correlation energy \cite{B2PLYP,wB97X-2,PBE0-DH,PBE0-2,SCAN0-2} can be added to the KS-DFT energy in order to improve 
the description of noncovalent interactions (e.g., vdW interactions). Alternatively, KS-DFT may also be combined with symmetry-adapted perturbation theory 
(SAPT) \cite{SAPT,SAPT1,SAPT2,SAPT3,SAPT4,SAPT5,SAPT6,SAPT7,SAPT8} to yield accurate results for noncovalent 
interactions \cite{SAPT_DFT,SAPT_DFT1,SAPT_DFT2,SAPT_DFT3,SAPT_DFT4,SAPT_DFT5,SAPT_DFT6}. In addition, to properly describe strong static correlation, it could be essential to develop a combined LC hybrid 
scheme with random phase approximations (RPAs) \cite{OD,Perdew09,RPA_Yang,RPA_Yang2} for small- to medium-sized systems or with thermally-assisted-occupation density functional theory 
(TAO-DFT) \cite{TAO-DFT,TAO-DFT2,TAO-DFT3} for large-sized systems.

\section{Conclusions} 

In conclusion, we have developed a comprehensive understanding of the physics involved in controlling the vdW interactions in rare-gas dimers. Specifically, we have examined the potential energy curves of the rare-gas dimers 
associated with a variety of interparticle interactions, using the highly accurate CCSD(T) method as well as other computational methods. The long-range interparticle interactions are shown to be essential for retaining the main 
features of conventional vdW interactions, which cannot be properly described by LDA, PBE, and possibly other semilocal density functionals in KS-DFT, but can be accurately described by MP2, CCSD, and possibly other fully 
nonlocal XC energy components from {\it ab initio} wavefunction methods. On the other hand, the nature of vdW interactions is shown to change drastically with the short-range interparticle interactions, wherein LDA, PBE, and 
possibly other semilocal density functionals in KS-DFT perform reasonably well for sufficiently short-range interparticle interactions (e.g., $\text{erfc}(\omega r)/r$ with $\omega$ = 0.30 bohr$^{-1}$ or larger). Therefore, our findings 
also support the main feature of the LC hybrid functionals for systems with Coulomb interactions. Although only the vdW interactions in rare-gas dimers and the triplet H$_2$ molecule are studied and discussed in this work, our 
conclusion may remain appropriate for other vdW-dominated systems.

\begin{acknowledgments} 

This work was supported by the Ministry of Science and Technology of Taiwan (Grant No.\ MOST104-2628-M-002-011-MY3), National Taiwan University (Grant No.\ NTU-CDP-104R7818), 
the Center for Quantum Science and Engineering at NTU (Subproject Nos.:\ NTU-ERP-104R891401 and NTU-ERP-104R891403), and the National Center for Theoretical Sciences of Taiwan. 
We would like to thank Prof.\ Peter Gill (ANU) and Su-Kuan Chu (NTU) for useful discussions. Yu-Ting Chen would like to give special thanks to her family. 

\end{acknowledgments}

\newpage 
\section*{Appendix: Asymptote of the interaction energy curve between two well-separated rare-gas atoms associated with the long-range (erf) interparticle interactions} 

Similar to the derivation for the Coulomb case (e.g., see Chapter 3 of Ref.\ \cite{vdW2}), we derive an analytical expression for the asymptote of the interaction energy curve between two well-separated rare-gas atoms associated with 
the long-range interparticle (nuclear-electron, electron-electron, and nuclear-nuclear) interaction operator $f(r)$: $\text{erf}(\omega r)/r$ (the erf interaction), based on a second-order perturbation theory \cite{PT}. Since in the Coulomb 
case, a rare-gas atom has no permanent multipole moments in its nondegenerate ground state \cite{MM}, presumably this remains correct for the erf interaction with a sufficiently large $\omega$ or for the erfc interaction 
[$f(r)$: $\text{erfc}(\omega r)/r$] with a sufficiently small $\omega$. Note also that the finite speed of propagation of electromagnetic signals is not taken into account in our derivation \cite{vdW2}. For brevity, the Einstein summation 
convention \cite{ESC} is adopted here. Based on this convention, when an index variable appears twice in a term, it implies a summation of that term over all possible values of the index. 

Consider a rare-gas atom $A$, composed of a nucleus situated at $\mathbf{r}_{\alpha=0}$ and $N_{A}$ electrons situated at $\mathbf{r_{\alpha}}$ ($\alpha = 1, 2, ..., N_{A}$) with respect to the nucleus of $A$. 
The electric potential at a point $\mathbf{r}$, due to the charge distribution, is 
\begin{equation} 
V_{A}(\mathbf{r}) = \frac{1}{4 \pi \epsilon_0} \sum\limits_{\alpha=0}^{N_{A}}e^{A}_{\alpha} f(|\mathbf{r}-\mathbf{r_{\alpha}}|), 
\label{eq:VA} 
\end{equation} 
where $e^{A}_{\alpha=0} = N_{A}e$ is the nuclear charge of $A$, and $e^{A}_{\alpha} = -e$ ($\alpha = 1, 2, ..., N_{A}$) is the charge of an electron. 
The Taylor series expansion of $V_{A}(\mathbf{r})$ around the nucleus of $A$ gives 
\begin{equation} 
\begin{split} 
V_{A}(\mathbf{r}) 
& = \frac{1}{4 \pi \epsilon_0} \bigg[\sum\limits_{\alpha}e^{A}_{\alpha}f(r) - \sum\limits_{\alpha}e^{A}_{\alpha}r_{i\alpha}\nabla_{i}f(r) + \frac{1}{2!}\sum\limits_{\alpha}e^{A}_{\alpha}r_{i\alpha}r_{j\alpha}\nabla_{i}\nabla_{j}f(r) 
+ \cdots \bigg] \\ 
& = \frac{1}{4 \pi \epsilon_0} \bigg[e^{A}_{tot} - \mu^{A}_{i}\nabla_{i} + Q^{A}_{ij}\nabla_{i}\nabla_{j} + \cdots \bigg]f(r), 
\end{split} 
\label{eq:VA2} 
\end{equation} 
where the first term is from an electric monopole $e^{A}_{tot}=\sum\limits_{\alpha}e^{A}_{\alpha}$, the second term is from an electric dipole, whose $i^{th}$ Cartesian component $\mu^{A}_{i}=\sum\limits_{\alpha}e^{A}_{\alpha}r_{i\alpha}$, 
the third term is from an electric quadrupole source, $Q^{A}_{ij}=\frac{1}{2!}\sum\limits_{\alpha}e^{A}_{\alpha}r_{i\alpha}r_{j\alpha}$, and so on. 

Consider a second rare-gas atom $B$, composed of a nucleus situated at $\mathbf{r}_{\beta=0}$ and $N_{B}$ electrons situated at $\mathbf{r_{\beta}}$ ($\beta = 1, 2, ..., N_{B}$) with respect to the nucleus of $B$. 
Let $\mathbf{R}$ be the separation distance vector pointing from the nucleus of $A$ towards the nucleus of $B$. The interaction energy between atoms $A$ and $B$ is 
\begin{equation} 
U_{AB} = \sum\limits_{\beta=0}^{N_{B}}e^{B}_{\beta}V_{A}(\mathbf{R}+\mathbf{r_{\beta}}), 
\label{eq:UAB} 
\end{equation} 
where $e^{B}_{\beta=0} = N_{B}e$ is the nuclear charge of $B$, and $e^{B}_{\beta} = -e$ ($\beta = 1, 2, ..., N_{B}$) is the charge of an electron. 
The Taylor series expansion of $V_{A}(\mathbf{R}+\mathbf{r_{\beta}})$ around the nucleus of $B$ gives 
\begin{equation} 
\begin{split} 
V_{A}(\mathbf{R}+\mathbf{r_{\beta}}) 
& = V_{A}(\mathbf{R}) + r_{i\beta}\nabla_{i}V_{A}(\mathbf{R}) + \frac{1}{2!}r_{i\beta}r_{j\beta}\nabla_{i}\nabla_{j}V_{A}(\mathbf{R}) + \cdots \\ 
& = \bigg[ 1 + r_{i\beta}\nabla_{i} + \frac{1}{2!}r_{i\beta}r_{j\beta}\nabla_{i}\nabla_{j} + \cdots \bigg]V_{A}(\mathbf{R}). 
\end{split} 
\label{eq:VA3} 
\end{equation} 
Substituting Eq.\ (\ref{eq:VA2}) and Eq.\ (\ref{eq:VA3}) into Eq.\ (\ref{eq:UAB}) produces 
\begin{equation} 
\begin{split} 
U_{AB} 
& = \sum\limits_{\beta}e^{B}_{\beta}\bigg[ 1 + r_{i\beta}\nabla_{i} + \frac{1}{2!}r_{i\beta}r_{j\beta}\nabla_{i}\nabla_{j} + \cdots \bigg]V_{A}(\mathbf{R}) \\ 
& = \bigg[e^{B}_{tot}+\mu^{B}_{i}\nabla_{i}+Q^{B}_{ij}\nabla_{i}\nabla_{j} + \cdots \bigg] \frac{1}{4\pi\epsilon_{0}} \bigg[e^{A}_{tot}-\mu^{A}_{i}\nabla_{i}+Q^{A}_{ij}\nabla_{i}\nabla_{j} + \cdots \bigg]f(R) \\ 
& = \frac{1}{4\pi\epsilon_{0}} \bigg[e^{A}_{tot}e^{B}_{tot} + \bigg(e^{A}_{tot}\mu^{B}_{i}\nabla_{i}-e^{B}_{tot}\mu^{A}_{i}\nabla_{i} \bigg)-\mu^{A}_{i}\mu^{B}_{j}\nabla_{i}\nabla_{j} + \bigg(e^{A}_{tot}Q^{B}_{ij}\nabla_{i}\nabla_{j} 
+ e^{B}_{tot}Q^{A}_{ij}\nabla_{i}\nabla_{j} \bigg) \\ 
&\ \ - \bigg(\mu^{A}_{i}Q^{B}_{jk}\nabla_{i}\nabla_{j}\nabla_{k} - \mu^{B}_{i}Q^{A}_{jk}\nabla_{i}\nabla_{j}\nabla_{k}\bigg) + Q^{A}_{ij}Q^{B}_{kl}\nabla_{i}\nabla_{j}\nabla_{k}\nabla_{l} + \cdots \bigg]f(R). 
\end{split} 
\label{eq:UAB2} 
\end{equation} 
Here $e^{B}_{tot}=\sum\limits_{\beta}e^{B}_{\beta}$, $\mu^{B}_{i}=\sum\limits_{\beta}e^{B}_{\beta}r_{i\beta}$, $Q^{B}_{ij}=\frac{1}{2!}\sum\limits_{\beta}e^{B}_{\beta}r_{i\beta}r_{j\beta}$, and so on. 
Since atoms $A$ and $B$ are both neutral, $e^{A}_{tot} = e^{B}_{tot} = 0$. Accordingly, 
\begin{equation} 
U_{AB} = \frac{1}{4\pi\epsilon_{0}} \bigg[-\mu^{A}_{i}\mu^{B}_{j}\nabla_{i}\nabla_{j} - \bigg(\mu^{A}_{i}Q^{B}_{jk} - \mu^{B}_{i}Q^{A}_{jk} \bigg)\nabla_{i}\nabla_{j}\nabla_{k} + Q^{A}_{ij}Q^{B}_{kl}\nabla_{i}\nabla_{j}\nabla_{k}\nabla_{l} 
+ \cdots \bigg]f(R) 
\label{eq:UAB3} 
\end{equation} 
can be expressed as a sum of dipole-dipole (dd), dipole-quadrupole (dq), quadrupole-quadrupole (qq), and other contributions. 

To evaluate the interaction energy between ground-state rare-gas atoms $A$ and $B$, the classical interaction energy given by Eq.\ (\ref{eq:UAB3}) should be first converted into quantum mechanical operator. Perturbation theory 
may then be adopted to obtain the various perturbation contributions to the interaction energy at large $R$. 

Let the Hamiltonian of an isolated rare-gas atom $X$ ($X$ = $A$, $B$) be $H^{X}$. The Schr\"odinger equation 
\begin{equation} 
H^{X} \psi^{X}_{n} = E^{X}_{n} \psi^{X}_{n} 
\label{eq:SEHX} 
\end{equation} 
is solved for the $n^{th}$ excited-state energy $E^{X}_{n}$ and wavefunction $\psi^{X}_{n}$, where the $n = 0$ case refers to the ground state. 
Accordingly, the full Hamiltonian of rare-gas atoms $A$ and $B$ can be expressed as 
\begin{equation} 
H = H^{A} + H^{B} + U_{AB}. 
\label{eq:H} 
\end{equation} 
The interaction energy between ground-state rare-gas atoms $A$ and $B$ can be calculated as 
\begin{equation} 
\Delta E_{int} = E_{0} - (E^{A}_{0} + E^{B}_{0}), 
\label{eq:Eint} 
\end{equation} 
where $E_{0}$ is the ground-state energy of $H$. 

To circumvent the need for solving the Schr\"odinger equation with Hamiltonian $H$, $E_{0}$ may be expressed in terms of $\{E^{A}_{n}, \psi^{A}_{n}; E^{B}_{n}, \psi^{B}_{n}\}$, based on perturbation theory \cite{vdW2,PT}. 
Since atoms $A$ and $B$ are well-separated, an appropriate unperturbed Hamiltonian is the sum of the Hamiltonians of the isolated atoms $A$ and $B$, 
\begin{equation} 
H_{0} = H^{A} + H^{B}. 
\label{eq:H0} 
\end{equation} 
Consequently, 
\begin{equation} 
H = H_{0} + U_{AB}, 
\label{eq:H2} 
\end{equation} 
where $U_{AB}$ given by Eq.\ (\ref{eq:UAB3}) is the perturbation. 

\subsection{Zeroth-Order Theory} 

$H_{0}\Psi^{(0)}_{n}=E^{(0)}_{n}\Psi^{(0)}_{n}$. 
At large $R$, the effects of electron exchange are insignificant. Accordingly, for the $n^{th}$ excited state, $\Psi^{(0)}_{n} = \psi^{A}_{r}\psi^{B}_{s}$ and $E^{(0)}_{n} = E^{A}_{r} + E^{B}_{s}$, where the isolated atoms $A$ and $B$ 
are described by quantum numbers $r$ and $s$, respectively. For the ground state, $\Psi^{(0)}_{0} = \psi^{A}_{0}\psi^{B}_{0}$ and $E^{(0)}_{0} = E^{A}_{0} + E^{B}_{0}$. Correspondingly, 
$\Delta E_{int} = E_{0} - (E^{A}_{0} + E^{B}_{0}) \approx (E^{(0)}_{0}) - (E^{A}_{0} + E^{B}_{0}) = (E^{A}_{0} + E^{B}_{0}) - (E^{A}_{0} + E^{B}_{0}) = 0$. 
Therefore, to obtain a nonvanishing $\Delta E_{int}$, it is necessary to go beyond the zeroth-order theory. 

\subsection{First-Order Theory} 

$E^{(1)}_{n}=\langle\Psi^{(0)}_{n}|U_{AB}|\Psi^{(0)}_{n}\rangle$. 
Since in the Coulomb case, the isolated rare-gas atom $X$ ($X$ = $A$, $B$) has no permanent multipole moments in its nondegenerate ground state \cite{MM}, presumably this holds true for the erf interaction with a sufficiently large 
$\omega$ or for the erfc interaction with a sufficiently small $\omega$. Accordingly, the dipole terms are vanished $\langle\psi^{X}_{0}|\mu^{X}_{i}|\psi^{X}_{0}\rangle=0$, the quadrupole terms are vanished 
$\langle\psi^{X}_{0}|Q^{X}_{ij}|\psi^{X}_{0}\rangle=0$, 
and so on. Therefore, the first-order correction to the ground-state energy is 
\begin{equation} 
\begin{split} 
E^{(1)}_{0} 
& = \langle\Psi^{(0)}_{0}|U_{AB}|\Psi^{(0)}_{0}\rangle \\
& = \langle\Psi^{(0)}_{0}|\frac{1}{4\pi\epsilon_{0}} \bigg[-\mu^{A}_{i}\mu^{B}_{j}\nabla_{i}\nabla_{j} - \bigg(\mu^{A}_{i}Q^{B}_{jk} - \mu^{B}_{i}Q^{A}_{jk} \bigg)\nabla_{i}\nabla_{j}\nabla_{k} + Q^{A}_{ij}Q^{B}_{kl}\nabla_{i}\nabla_{j}\nabla_{k}\nabla_{l} + \cdots \bigg]f(R)|\Psi^{(0)}_{0}\rangle \\
& = -\frac{1}{4\pi\epsilon_{0}}\bigg[\nabla_{i}\nabla_{j}f(R)\langle\Psi^{(0)}_{0}|\mu^{A}_{i}\mu^{B}_{j}|\Psi^{(0)}_{0}\rangle + \nabla_{i}\nabla_{j}\nabla_{k}f(R)\langle\Psi^{(0)}_{0}|\bigg(\mu^{A}_{i}Q^{B}_{jk} - \mu^{B}_{i}Q^{A}_{jk} \bigg)|\Psi^{(0)}_{0}\rangle \\
&\ \ - \nabla_{i}\nabla_{j}\nabla_{k}\nabla_{l}f(R)\langle\Psi^{(0)}_{0}|Q^{A}_{ij}Q^{B}_{kl}|\Psi^{(0)}_{0}\rangle + \cdots \bigg] \\ 
& = -\frac{1}{4\pi\epsilon_{0}}\bigg[\nabla_{i}\nabla_{j}f(R)\langle\psi^{A}_{0}\psi^{B}_{0}|\mu^{A}_{i}\mu^{B}_{j}|\psi^{A}_{0}\psi^{B}_{0}\rangle + \nabla_{i}\nabla_{j}\nabla_{k}f(R)\langle\psi^{A}_{0}\psi^{B}_{0}|\bigg(\mu^{A}_{i}Q^{B}_{jk} 
- \mu^{B}_{i}Q^{A}_{jk} \bigg)|\psi^{A}_{0}\psi^{B}_{0}\rangle \\
&\ \ - \nabla_{i}\nabla_{j}\nabla_{k}\nabla_{l}f(R)\langle\psi^{A}_{0}\psi^{B}_{0}|Q^{A}_{ij}Q^{B}_{kl}|\psi^{A}_{0}\psi^{B}_{0}\rangle + \cdots \bigg] \\
& = -\frac{1}{4\pi\epsilon_{0}}\bigg[\nabla_{i}\nabla_{j}f(R)\langle\psi^{A}_{0}|\mu^{A}_{i}|\psi^{A}_{0}\rangle\langle\psi^{B}_{0}|\mu^{B}_{j}|\psi^{B}_{0}\rangle \\
&\ \ + \nabla_{i}\nabla_{j}\nabla_{k}f(R) \bigg(\langle\psi^{A}_{0}|\mu^{A}_{i}|\psi^{A}_{0}\rangle\langle\psi^{B}_{0}|Q^{B}_{jk}|\psi^{B}_{0}\rangle - \langle\psi^{B}_{0}|\mu^{B}_{i}|\psi^{B}_{0}\rangle\langle\psi^{A}_{0}|Q^{A}_{jk}|\psi^{A}_{0}\rangle\bigg) \\ 
&\ \ - \nabla_{i}\nabla_{j}\nabla_{k}\nabla_{l}f(R)\langle\psi^{A}_{0}|Q^{A}_{ij}|\psi^{A}_{0}\rangle\langle\psi^{B}_{0}|Q^{B}_{kl}|\psi^{B}_{0}\rangle + \cdots \bigg] = 0. 
\end{split} 
\label{eq:E1} 
\end{equation} 
Accordingly, $\Delta E_{int} = E_{0} - (E^{A}_{0} + E^{B}_{0}) \approx (E^{(0)}_{0} + E^{(1)}_{0}) - (E^{A}_{0} + E^{B}_{0}) = E^{(1)}_{0} = 0$. 
Therefore, to obtain a nonvanishing $\Delta E_{int}$, it is also necessary to go beyond the first-order theory. 

\subsection{Second-Order Theory} 

$E^{(2)}_{n} = -\sum\limits_{m \neq n}\frac{|\langle\Psi^{(0)}_{m}|U_{AB}|\Psi^{(0)}_{n}\rangle|^{2}}{E^{(0)}_{m}-E^{(0)}_{n}}$. The second-order correction to the ground-state energy is 
\begin{equation} 
E^{(2)}_{0} = -\sum\limits_{m \neq 0}\frac{|\langle\Psi^{(0)}_{m}|U_{AB}|\Psi^{(0)}_{0}\rangle|^{2}}{E^{(0)}_{m}-E^{(0)}_{0}}, 
\label{eq:E2} 
\end{equation} 
which is always nonpositive. 

From Eq.\ (\ref{eq:UAB3}), if only the dipole-dipole contribution is retained, we have 
\begin{equation} 
U_{AB}^{dd} = -\frac{1}{4\pi\epsilon_{0}} \mu^{A}_{i}\mu^{B}_{j}\nabla_{i}\nabla_{j}f(R). 
\label{eq:UAB4} 
\end{equation} 
Accordingly, the second-order correction to the ground-state energy due to the dipole-dipole contribution is 
\begin{equation} 
\begin{split} 
E^{(2), dd}_{0} 
& = -\sum\limits_{m \neq 0}\frac{|\langle\Psi^{(0)}_{m}|U_{AB}^{dd}|\Psi^{(0)}_{0}\rangle|^{2}}{E^{(0)}_{m}-E^{(0)}_{0}} 
= -\sum\limits_{m \neq 0}\frac{|\langle\Psi^{(0)}_{m}|(-\frac{1}{4\pi\epsilon_{0}}) \mu^{A}_{i}\mu^{B}_{j}\nabla_{i}\nabla_{j}f(R)|\Psi^{(0)}_{0}\rangle|^{2}}{E^{(0)}_{m}-E^{(0)}_{0}} \\ 
& = -\frac{1}{(4\pi\epsilon_{0})^{2}}\sum\limits_{r\neq0}\sum\limits_{s\neq0}\frac{|\langle\psi^{A}_{r}\psi^{B}_{s}|\mu^{A}_{i}\mu^{B}_{j}\nabla_{i}\nabla_{j}f(R)|\psi^{A}_{0}\psi^{B}_{0}\rangle|^{2}}{E^{A}_{r} + E^{B}_{s} - E^{A}_{0} - E^{B}_{0}} \\ 
& = -\frac{1}{(4\pi\epsilon_{0})^{2}}\sum\limits_{r\neq0}\sum\limits_{s\neq0}\frac{\langle\psi^{A}_{r}\psi^{B}_{s}|\mu^{A}_{i}\mu^{B}_{j}\nabla_{i}\nabla_{j}f(R)|\psi^{A}_{0}\psi^{B}_{0}\rangle
\langle\psi^{A}_{0}\psi^{B}_{0}|\mu^{A}_{i'}\mu^{B}_{j'}\nabla_{i'}\nabla_{j'}f(R)|\psi^{A}_{r}\psi^{B}_{s}\rangle}{E^{A}_{r} + E^{B}_{s} - E^{A}_{0} - E^{B}_{0}} \\ 
& = -\frac{1}{(4\pi\epsilon_{0})^{2}}[\nabla_{i}\nabla_{j}f(R)][\nabla_{i'}\nabla_{j'}f(R)]\sum\limits_{r\neq0}\sum\limits_{s\neq0}\frac{\langle\psi^{A}_{r}\psi^{B}_{s}|\mu^{A}_{i}\mu^{B}_{j}|\psi^{A}_{0}\psi^{B}_{0}\rangle
\langle\psi^{A}_{0}\psi^{B}_{0}|\mu^{A}_{i'}\mu^{B}_{j'}|\psi^{A}_{r}\psi^{B}_{s}\rangle}{E^{A}_{r} + E^{B}_{s} - E^{A}_{0} - E^{B}_{0}} \\ 
& = -\frac{1}{(4\pi\epsilon_{0})^{2}}[\nabla_{i}\nabla_{j}f(R)][\nabla_{i'}\nabla_{j'}f(R)] \\ 
&\  \ \times \sum\limits_{r\neq0}\sum\limits_{s\neq0}\frac{\langle\psi^{A}_{r}|\mu^{A}_{i}|\psi^{A}_{0}\rangle\langle\psi^{B}_{s}|\mu^{B}_{j}|\psi^{B}_{0}\rangle\langle\psi^{A}_{0}|\mu^{A}_{i'}|\psi^{A}_{r}\rangle\langle\psi^{B}_{0}|\mu^{B}_{j'}|\psi^{B}_{s}\rangle}
{E^{A}_{r} + E^{B}_{s} - E^{A}_{0} - E^{B}_{0}}. 
\end{split} 
\label{eq:E2dd} 
\end{equation} 
In Eq.\ (\ref{eq:E2dd}), the ($r=0, s\neq0$) and ($r\neq0, s=0$) terms are excluded in the summation, due to the vanishing dipole terms, i.e., $\langle\psi^{X}_{0}|\mu^{X}_{i}|\psi^{X}_{0}\rangle = 0$ ($X$ = $A$, $B$). 

\begin{itemize} 

\item For the erf interaction, $f(R)=\frac{\erf(\omega R)}{R}$. 
\begin{equation} 
\begin{split} 
\nabla_{i}\nabla_{j}f(R) 
& = \nabla_{i}\nabla_{j}\frac{\erf(\omega R)}{R} = \nabla_{i}\bigg[\frac{1}{R}\nabla_{j}\erf(\omega R)+\erf(\omega R)\nabla_{j}\frac{1}{R}\bigg] \\ 
& = \nabla_{i}\bigg[\frac{1}{R}\frac{\partial\erf(\omega R)}{\partial R}\hat{R}_{j}-\frac{\erf(\omega R)}{R^{2}}\hat{R}_{j}\bigg] = \nabla_{i}\bigg[\frac{1}{R^{2}}\frac{\partial\erf(\omega R)}{\partial R}R_{j}-\frac{\erf(\omega R)}{R^{3}}R_{j}\bigg] \\
& = \frac{\partial}{\partial R}\bigg[\frac{1}{R^2}\frac{\partial\erf(\omega R)}{\partial R}\bigg]\hat{R}_{i}R_{j}+\frac{1}{R^{2}}\frac{\partial\erf(\omega R)}{\partial R}\delta_{ij}
-\frac{\partial}{\partial R}\bigg[\frac{\erf(\omega R)}{R^{3}}\bigg]\hat{R}_{i}R_{j}-\frac{\erf(\omega R)}{R^{3}}\delta_{ij} \\ 
& = R\bigg\{\frac{\partial}{\partial R}\bigg[\frac{1}{R^2}\frac{\partial\erf(\omega R)}{\partial R}-\frac{\erf(\omega R)}{R^{3}}\bigg]\bigg\}\hat{R}_{i}\hat{R}_{j}+\bigg[\frac{1}{R^{2}}\frac{\partial\erf(\omega R)}{\partial R}-\frac{\erf(\omega R)}{R^{3}}\bigg]\delta_{ij} \\ 
& = \bigg[-\frac{4\omega}{\sqrt{\pi}}\frac{1}{R^{2}}e^{-\omega^{2}R^{2}}-\frac{4\omega^{3}}{\sqrt{\pi}}e^{-\omega^{2}R^{2}}+\frac{3}{R^{3}}\erf(\omega R)-\frac{1}{R^{2}}\frac{2\omega}{\sqrt{\pi}}e^{-\omega^{2}R^{2}}\bigg]\hat{R}_{i}\hat{R}_{j} \\ 
&\ \ + \bigg[\frac{2\omega}{\sqrt{\pi}}\frac{1}{R^{2}}e^{-\omega^{2}R^{2}}-\frac{\erf(\omega R)}{R^{3}}\bigg]\delta_{ij}.
\end{split} 
\label{eq:erf} 
\end{equation} 
Since $e^{-\omega^{2}R^{2}}$ decays faster than polynomials when $R$ is large, \\
$\nabla_{i}\nabla_{j}f(R) \approx -\frac{\erf(\omega R)}{R^{3}}(\delta_{ij}-3\hat{R}_{i}\hat{R}_{j})$ at large $R$. 
Accordingly, 
\begin{equation} 
\begin{split} 
E^{(2), dd}_{0} 
& \approx -\frac{1}{(4\pi\epsilon_{0})^{2}} \frac{[\erf(\omega R)]^2}{R^{6}}(\delta_{ij}-3\hat{R}_{i}\hat{R}_{j})(\delta_{i'j'}-3\hat{R}_{i'}\hat{R}_{j'}) \\ 
&\  \ \times \sum\limits_{r\neq0}\sum\limits_{s\neq0}\frac{\langle\psi^{A}_{r}|\mu^{A}_{i}|\psi^{A}_{0}\rangle\langle\psi^{B}_{s}|\mu^{B}_{j}|\psi^{B}_{0}\rangle\langle\psi^{A}_{0}|\mu^{A}_{i'}|\psi^{A}_{r}\rangle\langle\psi^{B}_{0}|\mu^{B}_{j'}|\psi^{B}_{s}\rangle}
{E^{A}_{r} + E^{B}_{s} - E^{A}_{0} - E^{B}_{0}}. 
\end{split} 
\label{eq:E2dd2} 
\end{equation} 

Similar to the Coulomb case (e.g., see Chapter 3 of Ref.\ \cite{vdW2}), we adopt 
the rotational average of $\langle\psi^{A}_{r}|\mu^{A}_{i}|\psi^{A}_{0}\rangle\langle\psi^{A}_{0}|\mu^{A}_{i'}|\psi^{A}_{r}\rangle$ = $\frac{1}{3} \delta_{ii'} |\langle\psi^{A}_{r}|\mu^{A}|\psi^{A}_{0}\rangle|^{2}$, and 
the rotational average of $\langle\psi^{B}_{s}|\mu^{B}_{j}|\psi^{B}_{0}\rangle\langle\psi^{B}_{0}|\mu^{B}_{j'}|\psi^{B}_{s}\rangle$ = $\frac{1}{3} \delta_{jj'} |\langle\psi^{B}_{s}|\mu^{B}|\psi^{B}_{0}\rangle|^{2}$, where 
$\mu^{A}=\sum\limits_{\alpha}e^{A}_{\alpha}\mathbf{r_{\alpha}}$ and $\mu^{B}=\sum\limits_{\beta}e^{B}_{\beta}\mathbf{r_{\beta}}$. Also, note that 
\begin{equation} 
\sum\limits_{i=1}^{3}\sum\limits_{j=1}^{3}\sum\limits_{i'=1}^{3}\sum\limits_{j'=1}^{3}(\delta_{ij}-3\hat{R}_{i}\hat{R}_{j})(\delta_{i'j'}-3\hat{R}_{i'}\hat{R}_{j'})\delta_{ii'}\delta_{jj'} = 
\sum\limits_{i=1}^{3}\sum\limits_{j=1}^{3}(\delta_{ij}-3\hat{R}_{i}\hat{R}_{j})^{2} = 6. 
\label{eq:erf2} 
\end{equation} 
Therefore, from Eq.\ (\ref{eq:E2dd2}), 
\begin{equation} 
E^{(2), dd}_{0} \approx 
-\frac{1}{24\pi^{2}\epsilon_{0}^{2}}\frac{[\erf(\omega R)]^2}{R^{6}}\sum\limits_{r\neq0}\sum\limits_{s\neq0}\frac{|\langle\psi^{A}_{r}|\mu^{A}|\psi^{A}_{0}\rangle|^{2}|\langle\psi^{B}_{s}|\mu^{B}|\psi^{B}_{0}\rangle|^{2}}{E^{A}_{r} + E^{B}_{s} - E^{A}_{0} - E^{B}_{0}}.
\label{eq:E2dd3} 
\end{equation} 

From Eq.\ (\ref{eq:UAB3}), retaining also the dipole-quadrupole, quadrupole-quadrupole, and other contributions will produce additional terms in Eq.\ (\ref{eq:E2}), involving $\nabla_{i}\nabla_{j}\nabla_{k}f(R)$, 
$\nabla_{i}\nabla_{j}\nabla_{k}\nabla_{l}f(R)$, and so on. For the erf interaction, it can be shown that $\nabla_{i}\nabla_{j}f(R)$ decays more slowly than $\nabla_{i}\nabla_{j}\nabla_{k}f(R)$, and $\nabla_{i}\nabla_{j}\nabla_{k}f(R)$ 
decays more slowly than $\nabla_{i}\nabla_{j}\nabla_{k}\nabla_{l}f(R)$, and so on. Accordingly, $E^{(2)}_{0} \approx E^{(2), dd}_{0}$ at large $R$. Therefore, in the second-order theory, the interaction energy between rare-gas atoms 
$A$ and $B$ at large $R$ is 
\begin{equation} 
\begin{split} 
\Delta E_{int} 
& = E_{0} - (E^{A}_{0} + E^{B}_{0}) \approx (E^{(0)}_{0} + E^{(1)}_{0} + E^{(2)}_{0}) - (E^{A}_{0} + E^{B}_{0}) = E^{(2)}_{0} \approx E^{(2), dd}_{0} \\ 
& \approx 
-\frac{1}{24\pi^{2}\epsilon_{0}^{2}}\frac{[\erf(\omega R)]^2}{R^{6}}\sum\limits_{r\neq0}\sum\limits_{s\neq0}\frac{|\langle\psi^{A}_{r}|\mu^{A}|\psi^{A}_{0}\rangle|^{2}|\langle\psi^{B}_{s}|\mu^{B}|\psi^{B}_{0}\rangle|^{2}}{E^{A}_{r} + E^{B}_{s} - E^{A}_{0} - E^{B}_{0}}, 
\end{split} 
\end{equation} 
which has the $[\text{erf}(\omega R)]^2 R^{-6}$ asymptote. 

\item For the erfc interaction, $f(R)=\frac{\erfc(\omega R)}{R}$. 

In the second-order theory, the interaction energy between rare-gas atoms $A$ and $B$, 
$\Delta E_{int} = E_{0} - (E^{A}_{0} + E^{B}_{0}) \approx E^{(2)}_{0} = -\sum\limits_{m\neq0}\frac{|\langle\Psi^{(0)}_{m}|U_{AB}|\Psi^{(0)}_{0}\rangle|^{2}}{E^{(0)}_{m}-E^{(0)}_{0}}$, is always nonpositive. 
Therefore, it is necessary to go beyond the second-order theory to describe the repulsive interaction energy at large $R$ (as discussed in our paper), which is, however, beyond the scope of our discussion here. 

\end{itemize}

\bibliographystyle{pccp}

\newpage 

\newpage 

\newpage 

\newpage 

\newpage 

\newpage 

\newpage 

\newpage 

\newpage 

\newpage 

\newpage 

\end{document}